\documentclass[12pt,preprint]{aastex}

\slugcomment{2010AJ...139.2525H}

\shorttitle{A new Catalog of Isolated Galaxies from SDSS DR5.}

\shortauthors{Hern\'andez-Toledo et al.}

\begin{document}

\title{The UNAM-KIAS Catalog of Isolated Galaxies.}
\author{H. M. Hern\'andez-Toledo \altaffilmark{1} J. A. V\'azquez-Mata \altaffilmark{1} 
L. A. Mart\'{\i}nez-V\'azquez\altaffilmark{1}
\affil{Instituto de Astronom\'{\i}a, Universidad Nacional Aut\'onoma de M\'exico, A.P. 70-264, 04510 
M\'exico D. F., M\'exico}
Yun-Young Choi \altaffilmark{2}
\affil{Astrophysical Research Center for the Structure and
Evolution of the Cosmos, Sejong University, Seoul 143-747}
and 
Changbom Park \altaffilmark{3}}
\affil{Korea Institute for Advanced Study.  Seoul 130-012, Republic of Korea}
 
\altaffiltext{1}{E-mail: hector@astroscu.unam.mx}
\altaffiltext{1}{E-mail: jvazquez@astroscu.unam.mx}
\altaffiltext{2}{E-mail: yychoi@kias.re.kr}
\altaffiltext{2}{E-mail: cbp@kias.re.kr}

\begin{abstract}

A new catalog of isolated galaxies from The Sloan Digital Sky Survey (DR5) is presented. 1520 isolated galaxies were 
found in 1.4 steradians of sky. The selection criteria in this so called UNAM-KIAS catalog was implemented from a variation on the 
criteria developed by Karachentseva 1973 including full redshift information. Through an image processing pipeline that takes advantage 
from the high resolution ($\sim$ 0.4 $\arcsec$/pix) and high dynamic range of the SDSS images,  a uniform $g$ band morphological 
classification  for all these galaxies is presented.  We identify 80\% (SaSm) spirals (50\% later than Sbc types) on one hand, 
and a scarce population of early-type E(6.5\%) and S0(8\%) galaxies amounting to 14.5\% on the other hand. This 
magnitude-limited catalog is $\sim 80\%$ complete at 16.5, 15.6, 15.0, 14.6 and 14.4 magnitudes in the $ugriz$ bands respectively.
Some representative physical properties including SDSS magnitudes and color distributions, color-color diagrams, absolute magnitude-color, 
and concentration-color diagrams as a function of morphological type are presented.

The UNAM-KIAS Morphological Atlas is also released along with this paper. For each galaxy of type later than Sa, a mosaic is presented 
that includes:  (1) a $g$-band logarithmic image, (2) a $g$ band filtered-enhanced image where a Gaussian kernel of various 
sizes was applied and (3) an RGB color image from the SDSS database. For E/S0/Sa galaxies, in addition to the images in (1), (2) and (3), 
plots of $r$ band surface brightness and  geometric profiles (ellipticity $\epsilon$, Position Angle $PA$ and $A_{4}/B_{4}$ coefficients 
of the Fourier series expansions of deviations of a pure ellipse) are provided  \footnote{The mosaic images are available via our on-line 
web-site \\ 
http://www.astroscu.unam.mx/$\sim$hector/KIAS/kias.html}.

The size of the sample, the redshift completeness, the availability of high quality multicolor photometric data and
detailed morphological and spectroscopic information  make the UNAM-KIAS catalog of isolated galaxies a suitable sample   
to address important issues such as: (1) comparative studies of environmental effects, (2)  constraining the currently competing 
scenarios of galaxy formation and evolution, (3) the nature and  evolution of elliptical and spiral galaxies in the field, (4) the 
spectral properties of a statistically significant number of isolated galaxies and their evolution as a function of redshift and (5) 
the fraction of AGN in isolated environments among other important topics. 

The optimization and estimation of new structural parameters as well as important information to complement the existing one in other 
wavelengths is being carried out.

\end{abstract}

\keywords{Galaxies: spiral --
          Galaxies: elliptical and lenticular, cD --
          Galaxies: structure --
          Galaxies: photometry --
          Galaxies: fundamental parameters }

\section{Introduction} \label{S1}

The large and homogeneous redshift and image surveys like the 2dFGRS (Colles et al. 2001), SDSS (York et al. 2000) and more recently 
the 6dFGRS (Jones et al. 2004; 2009) have helped us to extend our view of the local universe. Not only the small-scale distribution 
of galaxies is being revealed but equally important, now it is possible to make accurate measurements of the relationship between the various 
physical properties of galaxies and the local environment. Recently, Park et al. (2007, 2008) and Park \& Choi (2008) used the SDSS data 
and studied the environmental dependence of the observed morphology at large and small smoothing scales to address the question of 
whether galaxy morphology depends primarily on the large-scale environment in which the galaxy initially formed or on a smaller scale 
environment that may reflect the influence of later evolutionary effects such as galaxy-galaxy interactions.  Park et al. (2008) pointed out 
that galaxies statistically become more isolated if they recently merged and that at a fixed large-scale density, more isolated galaxies are
more likely to be recent merger products.

Galaxy formation models within the $\Lambda CDM$ paradigm are every time better tuned 
to fit the joint luminosity/color/morphology distribution of low-redshift galaxies (c.f. De Lucia et al. 2006). These models assume that bulge 
formation takes place during mergers and that a galaxy can grow a disk later on provided that it is fed by an appreciable cooling flow. Spheroid 
formation is extended by assuming that bulges can also grow from disc instabilities. This additional mechanism can change the relative fractions 
of morphological types in these models. In the simulations of  De Lucia et al.  (2006), the final fractions of ellipticals, spirals and lenticulars 
brighter than -18 in the $V$ band are  17\%, 65\% and 18\%, very close to the fractions of 13\%, 67\% and 20\% reported by Loveday (1996). If 
bulge growth is switched-off, the above fractions become 7\%, 84\% and 8\% respectively, very close to the reported fractions for isolated 
galaxies from the CIG catalog (Karachentseva 1973) that are in common to the SDSS (DR6) by Hern\'andez-Toledo et al. (2008). This could be 
evidencing the impact of secular processes in the properties of isolated galaxies.

Uniformly selected and observed samples of galaxies that have not suffered any interaction with another normal galaxy or with a group 
environment over a Hubble time are thus crucial for studying intrinsic and secular processes able to affect  the structure, morphology and 
dynamics of galaxies, for instance, the formation and evolution of bars, rings, lopsidedness, and bulges.  Homogeneous observational data for 
such isolated galaxies are crucial for obtaining transparent scaling relationships and correlations that can be appropriately confronted to 
model predictions (see, e.g. Zavala et al. 2003; Avila-Reese et al. 2008).  It is worth to mention that an important requirement for all these 
samples is that of a well defined and strong isolation criteria, along with a uniform quality data acquirement in several wavelengths, and of 
statistical completeness.

In order to overcome some limitations of one of the most representative census of isolated galaxies in the local Universe started a few decades 
back by Karachentseva (1973) (e.g. lack of uniform, deep and high resolution CCD imaging, scarce of uniform and reliable photometric data as 
well as incompleteness in high resolution spectroscopic data),  we compiled a new catalog of isolated galaxies  from The Sloan Digital Sky 
Survey (DR5) as part of the IAUNAM-KIAS collaboration. This hereafter called UNAM-KIAS catalog has been compiled  by implementing 
a variation on the criteria developed by Karachentseva, including full redshift information as well as more accurate photometric 
information now available in the SDSS database. Using the three dimensional information, we are able to avoid missing galaxies
that happen to lie close to background or foreground galaxies. One principal aim of this new catalog is, providing new isolated galaxy 
candidates with the highest possible statistical and redshift completeness to complement the existing lists of similarly selected galaxies 
in the local Universe. Another relevant aim is to provide a body of observational parameters that will be used for a comparison analysis 
of the predicted properties of different models of galaxy formation and evolution.

In a companion paper by Hern\'andez-Toledo et al. 2008 (hereafter Paper 1) the results of a detailed morphological reevaluation of all  the 
available isolated CIG galaxies (549) from the Karachentseva (1973) catalog in common to the SDSS (DR6) was presented. The methodology was based
on an image processing scheme that takes advantage of the uniformity, photometric conditions and high resolution  (0.4 $\arcsec$/pix) of the 
SDSS CCD images.  Here we have applied the same image procedures to proceed with a morphological classification for the isolated 
galaxies of the UNAM-KIAS catalog.  A good understanding of these isolated, non--perturbed, galaxies and in particular of their morphological 
content is at the basis of any further correlations with other physical parameters and is one of the main legacies in the presentation of the 
UNAM-KIAS catalog.

The outline of the paper is as follows. Section 2 summarizes the relevant SDSS information and selection criteria used to compile the UNAM-KIAS 
catalog of isolated galaxies. In Section 3 the results of a detailed morphological classification are presented along with the public release of 
the UNAM-KIAS morphological atlas. In Section 4 some general properties, namely, completeness, apparent magnitude, color distributions, 
color-color diagrams, absolute magnitude-color and concentration-color diagrams as a function of morphological type are presented. Finally, in 
Section 5 a brief summary of the principal results and the main conclusions achieved are presented. The corresponding photometric and 
spectroscopic catalogs will be presented elsewhere.  The widely accepted cosmology with $H_o=70$ km s$^{-1}$ Mpc$^{-1}$, $\Omega_{m}=0.27$ and $\Omega_{\Lambda}=0.73$, suggested by the WMAP data (Spergel et al. 2003), is adopted throughout this paper.

\section{The Sample and Selection Criteria} \label{S2}

\subsection{The SDSS DR5 Data.}

The SDSS (York et al. 2000; Stoughton et al. 2002) is a survey to explore the large-scale distribution of galaxies and quasars by using 
a dedicated 2.5 m telescope at Apache Point Observatory (Gunn et al. 2006). The photometric survey has imaged roughly  $\pi$ steradians 
of the northern Galactic cap in five photometric band-passes denoted $u$, $g$, $r$, $i$, and $z$ centered at 3551, 4686, 6165, 7481, and 8931, 
respectively, by an imaging camera with 54 CCDs (Fukugita et al. 1996; Gunn et al. 1998). The limiting magnitudes of photometry at a 
signal-to-noise ratio of 5:1 are 22.0, 22.2, 22.2, 21.3, and 20.5 in the five band-passes, respectively. The median width of the 
point-spread function (PSF) is typically 1.4 $\arcsec$, and the photometric uncertainties are 2\% rms. After image processing 
(Lupton et al. 2001; Stoughton et al. 2002; Pier et al. 2003) and calibration (Hogg et al. 2001; Smith et al. 2002; Ivezic et al. 
2004; Tucker et al. 2006), targets are selected for spectroscopic follow-up observation.

We used a large-scale structure sample, DR4plus, from the New York University Value-Added Galaxy Catalog (NYU-VAGC; Blanton et al. 2005). 
The NYU-VAGC is a local galaxy catalog containing the SDSS galaxies with redshifts below about 0.3, and selects galaxies in the way similar
to that by which the SDSS Main Galaxy sample is made (see Strauss et al. 2002 for the detailed selection criteria for the SDSS Main galaxies).
The SDSS produced two galaxy samples; one is a flux-limited sample to extinction corrected apparent Petrosian $r$-band magnitudes of 17.77
(the Main Galaxy sample), and a color-selected and flux-limited sample extending to $r=19.5$ (the Luminous Red Galaxy sample).
The sample DR4plus is close to the SDSS Data Release 5 (Adelman-McCarthy et al. 2007). From this catalog, we selected the galaxies with 
$r$-band magnitudes in the range $14.5 \leq r_{\rm Pet} < 17.6$. We did not use galaxies in the southern Galactic Cap because of the narrow 
angular extent of the survey regions. Our survey region covers 4464 deg$^2$, containing 312,338 galaxies. Of the sample, approximately 
6\% lack measured redshifts because of fiber collisions, are assigned the spectroscopic redshift of the nearest neighbor, and are kept 
in our parent galaxy catalog. The isolated galaxies in our catalog are restricted to those with measured redshifts. However, we included 
galaxies with borrowed redshifts when the neighbors are searched to determine isolation of the target galaxies. 
This primary sample effectively has a magnitude range of only about 3.1 mag, which significantly limits 
the number of candidate isolated galaxies. To extend the range of magnitude, we attempt to include
the bright galaxies with  $r<14.5$. However, the spectroscopic sample of the SDSS galaxies is not 
complete for $r_{\rm Pet} < 14.5$. We thus searched for the various literature to borrow redshifts of 
the bright galaxies without SDSS spectra to increase the spectroscopic completeness. 
We added 5195 bright galaxies within our survey boundary to the primary sample (see Choi et al. 2007 for further details). 
In total, the final data set consists of 317,533 galaxies with known redshift and SDSS photometry.    

\subsection{Selection Criteria}

Our isolation criteria is specified by three parameters. The first is the extinction-corrected Petrosian $r-$band apparent magnitude 
difference between a candidate galaxy and any neighboring galaxy, $\Delta m_r$. The second is the projected separation to the neighbor 
across the line of sight, $\Delta d$. The third is the radial velocity difference, $\Delta V$.  Suppose a galaxy $i$ has a magnitude 
$m_{r,i}$ and $i$-band Petrosian radius $R_i$. It is regarded as isolated respect to potential perturbers if the separation $\Delta d$ 
between this galaxy and a neighboring galaxy $j$ with magnitude $m_{r,j}$ and radius $R_j$ satisfies the conditions

\begin{eqnarray}
 \Delta d \ge 100\times R_j   \\
  or \rm
 \Delta V \ge 1000 \rm km s^{-1}, 
 \end{eqnarray}

  or  the conditions

\begin{eqnarray}
 \Delta d < 100 \times R_j  \\
 \Delta V < 1000 \rm km s^{-1}  \\
 m_{r,j} \ge m_{r,i} + \Delta m_r, 
 \end{eqnarray}

for all neighboring galaxies. Here $R_{j}$ is the seeing-corrected Petrosian radius of galaxy $j$, measured in $i-$band using elliptical 
annuli to consider flattening or inclination of galaxies (Choi et al. 2007). We choose  $\Delta m_r = 2.5$. Using these criteria, we found a 
total of 1548 isolated galaxy candidates. Note that a magnitude difference of 2.5 in our selection criteria translates into a factor of about 
10 in brightness similar to that imposed by Karachentseva (1973). We removed spurious objects due to poor image deblending, bright stars 
within the area of a galaxy, images with strong diffraction spikes and the presence of strong diffuse light from other sources. We also 
removed galaxies fainter than $m_{r} = 15.2$ for reliability of our isolated galaxy sample because we require $\Delta m_{r} = 2.5$ to 
galaxies in the primary sample defined by the magnitude limit of $m_{r} = 17.6$. Thus the final number of isolated galaxies that will 
comprise the UNAM-KIAS catalog amounts to 1520.

Since the SDSS project scanned a quarter of the sky, one can figure out the number of CIG galaxies intersected by the  
Sloan survey, namely; (1050/2) $\sim$ 525 galaxies, similar to the number of CIG galaxies ($\sim$ 550) found in common to the SDSS 
(DR6) by Hern\'andez-Toledo et al. (2008). The CIG and the UNAM-KIAS catalogs have an interesting overlapping at velocities up to  
$<$ 10000 km s$^{-1}$ sharing 107 galaxies in common (see Table 1). This means that more than 400 CIGs were 
rejected after applying our selection criteria. Part of this can be explained by invoking equations 1 and 2 and imposing, to the sake of clarity, 
the condition $R_i = R_j$ (or $a_i = a_j$ in Karachentseva criteria). The isolation from similar-sized neighbors as projected in the sky 
is stronger in the UNAM-KIAS, rejecting cases where strict isolation is not satisfied by CIG galaxies and cases where isolation is satisfied 
but not equation 2 above. Equations (3, 4 and 5) also impose a stronger restriction to the presence of small-sized galaxies in the neighborhood 
of an isolated galaxy, rejecting a fraction of CIGs in such circumstances. Evidence for this comes from the refinement of the CIG 
isolation criteria in terms of collective tidal effects of small-sized neighbors (Verley et al. 2007) showing that an additional $\sim 20\%$ 
of the CIGs should be eliminated from the original sample. From the 107 CIG galaxies in common to the UNAM-KIAS, 77 belong to the refined 
sample in Verley et al (2007) indicating that our selection criteria includes much of such a refinement.

The UNAM-KIAS not only adds a significant number of new isolated galaxies at $v <$ 10000 km s$^{-1}$ vindicating some previously presumed 
non-isolated cases lacking radial velocity information (c.f Hern\'andez-Toledo et al. 2009), but becomes a natural extension of 
the CIG for $v >$ 10000 km s$^{-1}$ up to a bit more than 20000 km s$^{-1}$. It is therefore hoped that any joint study of the properties of 
these isolated galaxies should add necessarily more statistical significance and deeper insight to the origin of such galaxies.

\subsection{The UNAM-KIAS Catalog}

In Table 1 we list the general properties of the 1520 isolated galaxies sorted in Right Ascension: Column (1) gives a running identification 
number; Column (2), the galaxy name (following the IAU-designated SDSS naming convention), Column (3), an alternative name 
when available in the HyperLeda database, 
or CIG number when a match is produced, Column (4) the $\it{r}$-band Petrosian magnitude corrected for Galactic extinction according to 
Schlegel et al. (1998) reddening maps, Column (5) the $r$ band Absolute Magnitude (as described in the following section), (6) the galaxy 
redshift, Column (7) the $i$ band minor-to-major isophotal axis ratio, as an indicator of the apparent inclination, Column (8) the 
morphological type according to HyperLeda and finally Column (9) the morphological type according to this work.

\placetable{tbl-1}

\section{Morphological Content} \label{S3}
\subsection{Morphological Considerations}

The general criteria to carry out a morphological classification of the SDSS galaxies have been established and presented in 
a companion paper by Hernandez-Toledo et al. (2008) where $\sim$ 550 isolated galaxies from the CIG in common to the SDSS (DR6) database were 
classified in detail  following an image processing scheme that take advantage of the improved scale and dynamic range of the SDSS images 
and that enhances various low/high spatial frequency structural components in galaxies. Here only a brief description of our morphological 
considerations is presented. 

In order to discuss the optical morphology and its relationship to the global photometric properties, the isolated galaxies in the UNAM-KIAS catalog
were visually inspected through mosaics of images including, from upper-left to lower-right panel:  (1) a gray scale $g$-band image 
where a logarithmic transformation was applied to look for bright/faint internal/external details, (2) a $g$ band filtered-enhanced image 
where a Gaussian kernel of various sizes is applied to look for internal structure in the form of star forming regions, bars, rings and/or 
structure embedded into dusty regions. The filtered-enhancing techniques (Sofue 1993) allow the subtraction of the diffuse background in a 
convenient way to discuss different morphological details, including low surface brightness features, (3) an RGB color image from the SDSS 
database to complement our morphological analysis. These images are useful to visualize the spatial distribution of various morphological 
components in the images (blue colors are for recent SF and red colors for older populations/dusty components).
  
For E/S0 candidates, we further include in each mosaic, plots of surface brightness profiles and the corresponding geometric profiles 
(ellipticity $\epsilon$, Position Angle $PA$ and $A_{4}/B_{4}$ coefficients of the Fourier series expansions of deviations of a pure ellipse) 
from the $r$ band images to provide further evidence of boxyness/diskyness and other structural details. 

The classification of the sample followed the basic Hubble sequence. For spiral types the bulge to disk ratio as judged from the observed 
prominence of the bulge, tightness of the arms, and the degree of resolution of structure along the arms/outer disk  were considered. In the 
majority of the UNAM-KIAS spirals  these features are well recognized, however in some cases the presence of structures like dust lanes, 
prominent knots and the apparent tightening of the arms in the central regions may confuse the identification of structures like inner rings 
or bars. Outer rings/pseudo-rings (Buta 1995) were also identified when possible.  While the presence/absence of a bar was confirmed in 
some cases (SB), in the suspected cases we adopted the (SAB) nomenclature convention. The data presented in Table 1 reports the $i$ band 
minor-to-major isophotal axis ratio as an indicator of the inclination. This is important to evaluate the reliability of the classification  
presented here.

Figure 1 illustrates our image procedures, and shows 4 examples of UNAM-KIAS spirals classified according to the adopted criteria. The left panel 
shows a logarithmically scaled $g$ band image. The middle panel shows the corresponding filtered-enhanced version and  the right panel the 
corresponding $RGB$ image from the SDSS database. The relative importance of the bulge, arms and their degree of resolution into fragmented 
clumps in Sb, Sbc, Sc and Scd types is considered. Some main structural features like bars and rings were identified and sometimes suspected.  
Each galaxy is identified by its ISO number and the corresponding morphological type.  The average radial velocity of the UNAM-KIAS spirals is 
$\sim$ 10000 km s$^{-1}$ and in some cases the structural details are difficult to recognize at v $>$ 10000 km s$^{-1}$.
 
\placefigure{fig-1}

For early-type (E/S0) candidates, in addition to a careful inspection to the corresponding mosaic images, an evaluation of 
the geometric profiles after an isophotal analysis was carried out. Although the absolute value of the $A_{4}$ parameter depends on 
the inclination of the galaxy to the line of sight, its sign is useful to detect subtle disky features in the early-type candidates.  
A galaxy was judged to be an elliptical if the $A_{4}$ parameter showed: 1) no significant  boxy ($A_{4} < 0$) or disky ($A_{4} > 0$) 
trend in the outer parts, or 2) a generally boxy ($A_{4} < 0$) character in the outer parts. We further inspected the surface brightness 
profile for the presence/absence of 3) a linear component in the surface brightness-radius diagram. Central diskyness is considered not 
enough for an S0 classification. Figure 2 shows ISO 1404 to illustrate an elliptical galaxy and Figure 3 shows ISO 459 to illustrate a 
lenticular galaxy both at $v < 10000 km s^{-1}$.

\placefigure{fig-2}

\placefigure{fig-3}

The $g$-band filtered-enhanced image of ISO 1404 shows some sort of faint external envelope. We caution 
the reader about the reality of these features in some of our E/S0 candidates and that care must be taken about its 
interpretation, specially at velocities $v > 10000 km s^{-1}$ where our image methodology might not be good enough to resolve 
some structural details.

Another relevant goal of our image procedure is to isolate as much as possible the differences among lenticulars and very early-type 
spirals. Only from a more uniform and deeper survey like the CCD SDSS it is possible to systematically search for fainter 
features that could point to a more definite morphological classification. Here we use in addition, plots of both 
$\epsilon$ and $PA$ radial profiles as an auxiliary tool to disentangle among S0 or Sa cases. Significant and not necessarily 
coupled changes in $\epsilon$ and $PA$ radial profiles should be evidencing the presence of additional structure (in the form of a disk) for 
an S0 galaxy or in the form of arms, outer rings or envelopes for Sa galaxies. If further additional image processing did not show 
definite evidence of those features, we kept the galaxy type as lenticular. Figure 4 shows ISO 283 to illustrate a very early-type Sa galaxy. 
The advantages of the higher resolution and depth of the CCD SDSS images, in combination with our image procedures, is illustrated here 
for a galaxy with  $v < 10000 km s^{-1}$.
Nevertheless, the reader should be cautious about the classification for distant E/S0/Sa galaxies where it may be difficult to discriminate 
among representative structural details.

\placefigure{fig-4}

\subsection{Morphological Content: Results}

Table 2 reports our morphological evaluation for 1318 UNAM-KIAS galaxies, after removing 202 not classifiable galaxies (advanced mergers, 
compact, poorly defined objects, edge-on and highly inclined galaxies, as devised from their aspect ratio). The results are presented according 
to 3 velocity regimes: $v < 10000 km s^{-1}$, $v < 15000 km s^{-1}$ and $v < 20000 km s^{-1}$. Column (1) gives the morphological type, Column 
(2) the morphological code number following HyperLeda convention (code numbers 9 and 10 have been re-assigned to include Sdm and Sm galaxies), 
Columns (3) and (4) give the number $n$ of galaxies of each morphological type in each velocity regime and the corresponding fraction, respectively. 

\placetable{tbl-2}

Table 3 shows the results of our morphological classification this time reporting a morphological code number following the HyperLeda 
convention. Similarly to Table 2, code numbers 9 and 10 have been re-assigned to include Sdm and Sm galaxies.  Column (1) gives a running 
identification number; Column (2), the galaxy name (following the IAU-designated SDSS naming convention), Column (3) gives the morphological 
code number, Column (4) indicates the presence of a bar structure (absence = 0, suspected = 1 definite = 2), Column (5) indicates the presence 
of rings (absence = 0, inner = 1, outer =2, both = 3).

\placetable{tbl-3}

Figures 5, 6 and 7 illustrate the corresponding results in Table 2 according to 3 velocity regimes and in the form of histograms 
(upper panel) and cumulative distributions (lower panel). A classification avoiding transition E/S0 and S0/Sa cases was attempted as much 
as possible. For comparison, the results of our classification of 549 galaxies from the local $v < 10000 km s^{-1}$ CIG catalog in common 
to the SDSS (DR6) (Hern\'andez-Toledo et al. 2008) are also presented.

\placefigure{fig-5}
\placefigure{fig-6}
\placefigure{fig-7}

Figures 5-7 show that at $v < 10000 km s^{-1}$ our classification is consistent with that reported for the CIG galaxies in common to the 
SDSS database by Hern\'andez-Toledo et al. (2008) in the whole range of morphological types. The fraction of transition E/S0 and 
S0/Sa morphologies increases however at velocities $v > 10000 km s^{-1}$ denoting our increasing inability to distinguish structural 
details in early-type galaxies at such velocities. While the fraction of pure (E,S0) galaxies is $\sim$ 11\% at $v < 10000 km s^{-1}$, 
that fraction slightly increases up to $\sim$ 15\% at $v > 10000 km s^{-1}$, similar to the $\sim$ 14\% of (E,S0) galaxies reported by 
Sulentic et al. (2006) for the whole CIG catalog. For early-type Sa, Sab spirals, a fraction $\sim$ 14-15\% is obtained along the whole 
velocity range also consistent with the results in Hern\'andez-Toledo et al. (2008) but definitely at odds with the $\sim$ 6\% reported 
by Sulentic et al. (2006). Part of this discrepancy could be explained by the higher resolution and depth of the CCD SDSS images that in 
combination with our image procedures allow us to detect more easily high spatial frequency structure delineating arms, bars and rings.

According to our results, $\sim$ 80\% of the UNAM-KIAS galaxies are in the range of (Sa-Sm) types. While $\sim$ 30$\%$ of the spirals in this 
galaxy sample are earlier than Sbc, $\sim$ 50$\%$ are of Sbc type or latter (up to Sm types). Higher resolution is important to distinguish 
between inner rings and ring-like features produced by the tightening of the arms, the fragmentation degree of the arms in spirals or 
structures like bars, clumps, dust lanes, among others. Our filtering process enhances high spatial frequency structures in spirals 
(and sometimes, depending on the kernel size, in ellipticals too) making easier the distinction among various spiral types. However, it 
is also natural to expect that the filtering process loose power as the galaxy distance increase. This could explain the slight variations 
reported for the late-type fractions in Table 2 at the 3 velocity regimes. Care had to be taken of not over-interpreting high spatial 
frequency features during the classification and all these results should be considered as a first homogeneous insight into the 
morphological content of the isolated galaxies in the UNAM-KIAS sample.

The information concerning bars (confirmed and presumed) indicates that about 62.9$\%$ of the isolated galaxies in the UNAM-KIAS catalog 
show evidence of barred structure: for 26.3$\%$ the evidence is clear (SB galaxies) and for 36.6$\%$ the bars are weak or suspected (SAB 
galaxies). The bar fraction (SAB + SB) is 27.8$\%$ for early types and 35$\%$ in late types.  However we caution the reader about these 
numbers. The bar fraction reported here is mainly the result of our visual reevaluation of the mosaic images and only a fraction of the 
UNAM-KIAS spirals (mainly Sa types) were judged for the presence of bars through a photometric analysis of the $r-$band isophotal ellipticity 
and $PA$ profiles (c.f Wozniak et al. 1995). Furthermore, the fraction of bars could increase to an additional 10-20$\%$ if the corresponding 
analysis were extended into the NIR bands (c.f. Hern\'andez-Toledo et al. 2007; 2008). Similarly, the information for rings (inner, outer 
rings and pseudo-rings) in our sample is tentatively reported at the 36$\%$ level.

\subsection{The Atlas}

1420 isolated galaxies are presented in the form of mosaic images. Compact objects or other objects of non-definite nature were eliminated 
from the Atlas. For spiral galaxies of types later than Sa, we include, from upper-left to lower-right panels: 
(1) a gray scale $g$-band logarithmically transformed image 
(2) a $g$ band filtered-enhanced version of the image in 1) where a Gaussian kernel of variable size was applied and
(3) an RGB color image from the SDSS database.

For E/S0/Sa galaxies, in addition to the images in  (1), (2) and (3), plots of the surface brightness and geometric profiles (ellipticity 
$\epsilon$, Position Angle $PA$ and $A_{4}/B_{4}$ coefficients of the Fourier series expansions of deviations of a pure ellipse) from the 
$r$ band images are provided. 

The morphological diversity of the UNAM-KIAS sample is demonstrated in this Atlas. Interested people are invited to visit our 
on-line web-site \\
(http://www.astroscu.unam.mx/$\sim$hector/KIAS/kias.html).

\section{General Properties of the UNAM-KIAS Sample} \label{S4}

A first preliminary view of the relationship among some physical parameters of the galaxies in the UNAM-KIAS sample is presented.
We start by showing in Figure 8 an histogram of the redshift distribution for the original number of galaxies (1520) in the UNAM-KIAS 
catalog.  Plots are shown distinguishing by morphological type: early (ellipticals and lenticulars) and late (spirals and irregulars) 
types, based on our own classification. 
 
\placefigure{fig-8}

The mean redshift for the UNAM-KIAS sample is $< z > = 0.032$, which corresponds to a comoving distance of 143 $h^{-1}$ Mpc for an 
$\Omega_{m} = 0.27$;  $\Omega_{\Lambda} = 0.73$ cosmology. The redshift distribution of the spirals in the UNAM-KIAS catalog is   
consistent with a Gaussian distribution and does not show evidence of obvious concentrations associated with major components of large-scale 
structure. However, the bins centered at $z = 0.027$ and 0.042 in the E/S0 distribution show two peaks slightly deviating from a Gaussian. This  
may be interpreted as a departure from an homogeneous distribution but a more careful analysis needs to be done. The apparent homogeneity 
in redshift distribution of the galaxies in the UNAM-KIAS catalog points to a sample that is close to a local homogeneous component of the 
isolated galaxy distribution.

Aitoff projections in right ascension and declination in Figures 9 and 10 show the distribution of the UNAM-KIAS sample on the sky at 
3000 km s${-1}$ velocity intervals covering the velocity interval from 0 to 21000 km $s^{-1}$. Due to their small number, galaxies at
v $>$ 21000 km $s^{-1}$ were not considered. After a new search for redshifts of Abell cluster cores in the literature we show their position 
at every redshift range according to increasing richness classes from 0 (crosses), 1 (asterisks), 2 (rhombus) and 3 (triangles). 

\placefigure{fig-9}
\placefigure{fig-10}

There is no apparent association between the positions of the Abell cluster cores and the UNAM-KIAS galaxies in the 0 to 9000 km $s^{-1}$ 
velocity intervals. However as the velocity increases, the presence of Abell clusters is more significant and a possible 
association of some our galaxies to those structures may not be negligible. Any correspondence of our galaxies to complex local large-scale 
structure is something that needs to be further explored in more detail.

Figure 11 shows the distribution of isolated galaxies in the redshift $z$-Absolute Magnitude diagram. Red dots represent E/S0 
isolated galaxies while black dots correspond to spiral isolated galaxies. We use the seeing-corrected isophotal 
axis ratio in the $i$ band to take into account the inclination of the galaxies. Cross symbols indicate galaxies with inclinations greater 
than 70${\deg}$. The $r$-band absolute magnitude $M_ {r}$ is estimated from the SDSS apparent magnitudes by using the expression:
\begin{eqnarray}
m_{r} -M_{r} =  5 {\rm log} (r(1+z))+ 25 + K(z) + E(z),
\end{eqnarray}

where $K(z)$ is the $K$-correction and E(z) = 1.6(z-0.1) is the mean luminosity evolution correction (Tegmark et al. 2004), $r$ is the 
comoving distance corresponding to redshift $z$. We also omit the $+5{\rm log}h$ term in the absolute magnitude. We adopt a flat $\Lambda$CDM 
cosmology with density parameters $\Omega_{\Lambda} = 0.73$ and $\Omega_{m} = 0.27$ to convert redshift to comoving distance. The $r$-band 
rest-frame magnitude are Galactic extinction corrected (Schlegel et al. 1998) and $K$-corrected to redshift of 0.1 (Blanton et al. 2003). 
This makes galaxies at $z=0.1$ have $K$-correction of $-2.5 \log (1+0.1)$, independent of their spectral energy distributions (SEDs).

\placefigure{fig-11}
   
A wide range in absolute magnitudes is observed for the UNAM-KIAS sample.  Notice a scarcity of nearby dwarf isolated E/S0 galaxies, and that 
bright E/S0 galaxies are preferentially found more at intermediate to high redshifts than spirals. The covered redshift range of 
UNAM-KIAS sample has increased with respect to that in the CIG catalog by about a factor of two. However, although deeper, this is also a 
magnitude-limited sample.  
 
Figure 12 presents the distribution of apparent magnitudes in the $ugriz$ SDSS photometric bands. Galaxies are sorted into E/S0 (red)
and Sa-Sm (black). Apparent magnitudes were corrected for Galactic extinction according to Schlegel et al. (1998).
 
\placefigure{fig-12}

The cutoff at about $\sim 15.2$ mag in the $r$ band is a consequence of the selection criteria applied to this galaxies. The $g$ band 
magnitude distribution shows a similar range of variation as that shown in the $B$ band for the CIG galaxies in Karachentseva catalog, 
considering a proper photometric transformation.     

Figure 13 shows the ($u-r$), ($r-i$), ($g-r$) and ($i-z$) color distributions for the galaxies in the UNAM-KIAS catalog. Red histograms are 
denoted for E/S0 galaxies while black histograms are for Sa-Sm galaxies. The computed colors use extinction (Schlegel et al. 1998)  and $K$ 
corrected magnitudes (Blanton 2003).

\placefigure{fig-13}

This figure shows how the ($u-r$) color index is more effective to distinguishing among different galaxy types, although a significant overlap 
is still appreciated at redder colors (see e.g. Park \& Choi 2005). The ($u-r$) color could be used as a measure of 
star formation activity of galaxies in the recent past, as suggested by Choi et a. (2007).

Figure 14 presents the apparent ($u - g$) vs ($g - r$),  ($g - r$) vs ($r - i$),  ($r - i$) vs ($i - z$) and ($g - r$) vs ($u - r$) color-color 
diagrams for all the galaxies in the UNAM-KIAS  catalog.  Cross symbols (red) denote E/S0 galaxies and inverted triangles (black) represent 
Sa-Sm galaxies.  

\placefigure{fig-14}

Tighter color-color distributions are appreciated for the UNAM-KIAS Catalog compared to similar color-color diagrams for isolated galaxies 
from the SDSS(DR1) in Allam et al. 2005. 

Figure 15 presents the behavior of the UNAM-KIAS isolated galaxies in the $C_{in}$ vs ($u-r$) and $C_{in}$ vs 
Morphological type diagrams. The inverse concentration index, $C_{in}$ is defined as $R_{50}/R_{90}$ where $R_{50}$ and $R_{90}$ are the radii 
from the center of a galaxy containing 50$\%$ and 90$\%$ of the flux in the $i$ band. Red dots are for E galaxies, blue dots for S0 galaxies and green 
dots for Sa galaxies. 

\placefigure{fig-15}

Late-types are loosely distributed in the  $C_{in}$ vs ($u-r$) diagram showing a broad overlap with early-type galaxies. Notice how  highly inclined 
galaxies are shifted towards redder colors.  The lower panel of Figure 15 shows that the inverse concentration index follows a loose tendency with 
morphological type. Early types are the most concentrated.  The tendency is fuzzy due to the big scatter of $C_{in}$ values at later types. The high 
degree of overlap in morphological types at a given  $C_{in}$ value in this diagram illustrates about the difficulty to distinguish not only between E/S0, 
S0/Sa classes but also between early and late-types by simply establishing cuts in  the $C_{in}$ domain. Figure 15 compares with the similar plot for 
the randomly chosen SDSS galaxies brighter than $r=15.9$ shown in Figure 1 of Park \& Choi (2005).

Finally Figure 16 shows the $C_{in}$ vs $M_{r}$ and color magnitude ($u-r$) vs $M_{r}$ 
diagrams.

\placefigure{fig-16}

The inverse concentration index $C_{in}$ of early-type galaxies is nearly independent of absolute magnitude but a very slight dependence 
towards lower concentration is noticed at fainter magnitudes. For late-type galaxies it is more difficult to see a tendency due to the large scatter 
at all absolute magnitude intervals. This trend has been also observed for the general SDSS early-type galaxies (see Fig. 3 of Choi et al. 2007).
We inspect the color-magnitude diagram in the lower panel of Figure 16 to explore how a segregation into early (red symbols) and late-type 
(black symbols) sequences is possible in isolated environments. We notice a tendency of isolated early-type galaxies to have redder colors at 
brighter absolute magnitudes.  In the case of the general SDSS galaxies the red sequence of early-type galaxies has a break in the slope at about
$M_r = -19.6$ (Fig. 3 of Choi et al. 2007). Isolated late-type galaxies show a larger scatter in ($u-r$) color compared to early-type galaxies. Notice 
that the color distribution of late-type galaxies overlaps with early types practically at all absolute magnitude intervals and thus a line dividing 
the sequences of red and blue galaxies is difficult to find. Although the fraction of early types located outside the red sequence is low, this 
diagram illustrates the danger of generating samples of early-type galaxies by a simple cut in the absolute magnitude-color diagram. 
We also detect a relatively small fraction of blue early-type galaxies that may be associated with recent merger events. Notice that a 
significant fraction of the most inclined galaxies are located towards fainter magnitudes and redder colors due to internal extinction (see also 
Fig. 12 of Choi et al. 2007 for a comparison).

\subsection{Completeness}

An estimate of the statistical completeness of the sample by means the ($V/V_{m}$) test (Schmidt 1968) is presented. For each object we 
estimated the volume $V$ at a radius corresponding to its distance and the maximum volume $V_{m}$ at a radius corresponding to the 
maximum distance given the magnitude limit of the UNAM-KIAS sample. We then calculate the average $V/V_{m}$ for objects 
brighter than a given magnitude limit. Figure 17 shows the results of $<V/V_{m}>$ test as a function of limiting $ugriz$ apparent 
magnitudes. Approximately  1450 isolated galaxies were included in the test. Due to their small number, galaxies with radial velocities 
in excess of 20000 km s$^{-1}$ were eliminated from this analysis. 

\placefigure{fig-17}

The decreasing trend observed in the (12-13.6) $r$ band interval is interpreted as a statistical fluctuation or incompleteness due to small 
number of galaxies in those magnitude bins. Since the parent galaxy sample from which the UNAM-KIAS was select is intrinsically incomplete at 
bright magnitudes and in spite that bright galaxies from other catalogs were added, the effect this supplement had on completeness appeared to be 
non-significant and reflects the difficulty of having nearby isolated bright galaxies. At intermediate (13.7-15.2) $r$ band magnitudes where 
most of our galaxies yield statistical significance to the test, the UNAM-KIAS is about ~80
level of completeness for galaxies satisfying our selection criteria. Incompleteness of this sample beyond 15.2 $r$ band Petrosian 
magnitudes, is a natural consequence of our selection criteria that excludes galaxies beyond that magnitude range to ensure that we are within the 
completeness limit of the parent sample (Section 2.2). Notice how the faint-end completeness limit in the $r$ band magnitude is correctly 
recovered from this test.

\section{Summary and Conclusions} \label{S5}

In order to overcome some limitations of the previous major sample of isolated galaxies in the local universe by 
Karachentseva (1973); to mention, a lack of uniform, deep and high resolution CCD imaging, the scarcity of uniform and reliable photometry 
and high resolution spectroscopic incompleteness, we compiled a new catalog of isolated galaxies  from the Sloan Digital Sky Survey (DR5)  
that adds a significant number of new isolated galaxies in the local Universe also complementing the previous census.  
 
This catalog coined as UNAM-KIAS, has been compiled by implementing a variation on the criteria developed by Karachentseva, including full 
redshift information as well as more accurate photometric information now available in the SDSS database. 

We have taken advantage from the high resolution ($\sim$ 0.4 $\arcsec$/pix) and high dynamic range of the SDSS images by implementing 
a simple digital image processing scheme that enhances structural details in galaxies of various types supporting a detailed morphological 
characterization for the isolated galaxies in the UNAM-KIAS catalog. A Morphological Atlas containing detailed  mosaic images for the galaxies 
is also presented and released along with this paper. The classification presented here  preserves the optically observed morphology but takes 
into account structural information provided from the digital image processing, the isophotal 
analysis (E/S0/Sa cases) and the color information provided by the SDSS RGB images. 

The isolated galaxy candidates in the UNAM-KIAS catalog show a wide morphological diversity, from  E to Sm  types. 
80$\%$ of these galaxies are in the range of (Sa-Sm) types, 14.5$\%$ are of (E,S0) type and 50$\%$ are of Sbc type or latter (up to Sm types).  
Although this is an attempt  to provide an uniform morphological classification, the reader should notice that this classification is not free of uncertainties, 
specially for some E/S0/Sa galaxies  at larger distances where our image procedures  may no longer be enough for disentangling among 
their characteristic structural properties. A tentative fraction of about  62.9$\%$ of the isolated galaxies in the UNAM-KIAS catalog show evidence of barred 
structure: for 26.3$\%$ the evidence is clear (SB galaxies) and for 36.6$\%$ the bars are weak or suspected (SAB galaxies). The bar fraction (SAB + SB) by 
morphological type is 27.8$\%$ for early types and 35$\%$ for late types. 

Although we did not a systematic search for bars in the present study, we notice that the tentative fraction of bars (62.9$\%$) is not so 
different from that reported for galaxies in other environments. This could be suggesting that interactions and the global effects of the 
group/cluster environment are not crucial for the formation/destruction of bars.  However, in an analysis of a volume-limited sample of the 
general SDSS galaxies, Lee \& Park (2008) have found  a dependence of the bar fraction on the large-scale background density  when the bars 
are visually identified. It is fundamental thus to carefully infer the fraction of bars in the UNAM-KIAS sample by an 
homogeneous method.  

A first exploration of the general properties of the  galaxies in the UNAM-KIAS catalog and of the relations among some of their 
representative physical parameters leaded us to some important conclusions:

a) The spatial distribution of the UNAM-KIAS galaxies, as seen from their redshift distribution does not show evidence of obvious concentrations 
associated with major components of large-scale structure in the local Universe but a more careful study in this direction needs to be done.  This apparent 
homogeneity points to a sample that is close to a local homogeneous component of the isolated galaxy distribution. 

b) Although the UNAM-KIAS galaxies comprise a nearby sample, its redshift interval has increased with respect to that in the CIG catalog
 by about a factor of two, making it deeper and representative of the galaxies in this environment in the local Universe.

c) The UNAM-KIAS sample is also diverse in terms of  physical properties explored here. A scarcity of  nearby dwarf
isolated E/S0 galaxies and a  tendency of  bright E/S0 galaxies being farther away than spiral galaxies is observed.  

d) The UNAM-KIAS catalog of isolated galaxies is a magnitude-limited sample that is reasonably  complete ($\sim$ 80$\%$) up to 15.2 $r$ band 
apparent magnitudes, providing statistical significance to studies of these galaxies in isolated environments.

The finding of reasonably nearby isolated galaxies uniformly-selected and with detailed morphological information is of high relevance. The 
UNAM-KIAS catalog along with the most isolated galaxies of the CIG catalog that are in common to the SDSS will provide a unique database 
that can be used for several studies, including (i)  studies of the environmental effects on galaxies belonging to groups and clusters, and (ii)  for 
confronting with theoretical and model predictions of galaxy evolution.

Other relevant aims, derived from the present study, are to provide as much as possible a new body of observational parameters for these 
isolated galaxies in different wavelengths that will greatly enable to a major understanding of the nature of the galaxies in this environment.

\begin{acknowledgements}

HMHT and JAVM  acknowledge CONACYT through grant 42810.
CBP and YYC acknowledge the support of the Korea Science and Engineering
Foundation (KOSEF) through the Astrophysical Research Center for the
Structure and Evolution of the Cosmos (ARCSEC).

Funding for the SDSS and SDSS-II has been provided by the Alfred P. Sloan Foundation, 
the Participating Institutions, the National Science
Foundation, the U.S. Department of Energy, the National Aeronautics and
Space Administration, the Japanese Monbukagakusho, the Max Planck
Society, and the Higher Education Funding Council for England.
The SDSS Web Site is http://www.sdss.org/.
The SDSS is managed by the Astrophysical Research Consortium for the
Participating Institutions.

The Participating Institutions are the
American Museum of Natural History, Astrophysical Institute Potsdam,
University of Basel, Cambridge University, Case Western Reserve University,
University of Chicago, Drexel University, Fermilab, the Institute for
Advanced Study, the Japan Participation Group, Johns Hopkins University,
the Joint Institute for Nuclear Astrophysics, the Kavli Institute for
Particle Astrophysics and Cosmology, the Korean Scientist Group, the
Chinese Academy of Sciences (LAMOST), Los Alamos National Laboratory,
the Max-Planck-Institute for Astronomy (MPIA), the Max-Planck-Institute
for Astrophysics (MPA), New Mexico State University, Ohio State University,
University of Pittsburgh, University of Portsmouth, Princeton University,
the United States Naval Observatory, and the University of Washington.
This research has made use of the NASA/IPAC Extragalactic Database (NED) which is 
operated by the Jet Propulsion Laboratory, California Institute of Technology, 
under contract with the National Aeronautics and Space Administration. 
We acknowledge the usage of the HyperLeda database (http://leda.univ-lyon1.fr).

HMHT and JAVM  acknowledge annonymous referee for her/his careful reading and
appropriate questioning that greatly improved this manuscript. 

\end{acknowledgements}

\clearpage

\begin{deluxetable}{lllllcccll} 
\tablecolumns{10}
\tablewidth{0pc} 
\tablecaption{The UNAM-KIAS sample of Isolated Galaxies.
\label{tbl-1}}
\tablehead{
\colhead{} ID   & SDSS ID & A &  $\it{r}$  &  $M_r$  &  $\it{z}$  & $(b/a)_r$   & HLeda  &  Type  &}  
\startdata

1   &   {\small J072246.73+413929.6}     &             &   13.82   &   -20.23   &   0.023   &   0.39   &       Sb      &   SBbc(r)   &      \\
2   &   {\small J072333.23+412605.5}      &            &   13.41   &   -20.80   &   0.027   &   0.87   &       SABb      &   SBc(r)    &      \\
3   &   {\small J072635.39+431746.8}      &            &   12.77   &   -19.43   &   0.010   &   1   &       E      &   E   &      \\
4   &   {\small J072642.64+391724.5}      &            &   14.64   &   -20.61   &   0.045   &   0.54   &       S?      &   SABbc   &      \\
5   &   {\small J072719.37+442538.3}      &            &   14.85   &   -19.73   &   0.032   &   0.39   &       S?      &   Sa   &      \\
6   &   {\small J072954.29+372706.3}      &            &   13.45   &   -21.37   &   0.035   &   0.64   &       Sbc      &   SBc(r)   &      \\
7   &   {\small J073054.71+390110.1}      &            &   14.16   &   -19.19   &   0.019   &   0.3   &       Sb      &   SBc(r)   &      \\
8   &   {\small J073548.83+401938.2}      &            &   14.67   &   -20.54   &   0.042   &   0.68   &       S?      &   -   &      \\
9   &   {\small J073901.44+315452.9}      &            &   14.26   &   -20.85   &   0.040   &   0.77   &            &   SB0   &      \\
10   &  {\small J074001.42+321140.5}      &            &   14.65   &   -19.68   &   0.027   &   0.51   &       S?      &   Sab   &      \\
11   &  {\small J074018.11+282751.4}      &            &   14.31   &   -19.65   &   0.022   &   0.9   &            &   Sab   &      \\
12   &  {\small J074022.74+231629.9}      &            &   13.29   &   -18.17   &   0.007   &   0.66   &       E      &   E   &      \\
13   &  {\small J074109.47+492347.0}      &            &   14.66   &   -19.98   &   0.030   &   0.96   &       S?      &   SABc   &      \\
14   &  {\small J074112.48+424457.7}      &            &   13.47   &   -21.44   &   0.036   &   0.64   &       E?      &   Sa   &      \\
15   &  {\small J074127.73+294009.0}     &            &   14.55   &   -19.56   &   0.023   &   0.47   &            &   SABb(r)   &      \\
16   &  {\small J074158.62+231035.0}     &            &   14.18   &   -21.13   &   0.043   &   0.71   &       S?      &   E/S0   &      \\
17   &  {\small J074232.38+491127.9}      &           &   14.35   &   -17.79   &   0.009   &   0.45   &       Sa      &   Sd   &      \\
18   &  {\small J074252.09+220645.7}      &           &   13.81   &   -20.23   &   0.028   &   1   &            &   SBc(r)   &      \\
19   &  {\small J074330.38+225549.9}      &           &   14.52   &   -19.71   &   0.024   &   0.28   &       Sbc      &   Sc   &      \\
20   &  {\small J074403.33+330438.6}      &           &   14.64   &   -21.16   &   0.055   &   0.73   &       S?      &   SABbc   &      \\

\enddata

\end{deluxetable}

\clearpage 

\begin{deluxetable}{lcccccccc} 
\tablecolumns{9}
\tablewidth{0pc} 
\tablecaption{Results of the morphological classification for the  UNAM-KIAS sample at three velocity regimes.
\label{tbl-1}}
\tablehead{
\colhead{} & & & {\small v$<$10000km $s^{-1}$} &  & {\small v$<$15000km $s^{-1}$} &  & {\small v$<$20000km $s^{-1}$} \\
\cline{1-8} \\
\colhead{} Type &  T  & $n$ & $n/835$ & $n$ & $n/1203$ & $n$ & $n/1318$ }
\startdata 

E       &  -5   & 42   & 0.050   &   69 &  0.057  &  86      &  0.065   &  \\
E/S0  &  -3  & 9     & 0.011   &   37 &  0.031  &  45      &  0.034   &  \\
S0     &  -2   & 56   & 0.067  &   89  &  0.074  &  105   &  0.080    &  \\
S0/Sa &  0  & 12   & 0.014  &   32  &  0.027  &  39      &  0.030   &  \\
Sa     &  1    &  51  & 0.061  &   67  &  0.056  &  76      &  0.058   &  \\
Sab   &  2    &  73  & 0.087  &  115 &  0.096  &  126   &  0.096   &  \\
Sb     &  3    & 110 & 0.132  &  171 &  0.142  &   185  & 0.140    &  \\
Sbc   &  4    & 114 & 0.137  &  179 &  0.149  &  197   & 0.149    &  \\
Sc     &  6    & 215 & 0.257  &  283 &  0.235  &  297   & 0.225    &  \\
Scd   &  7    &  93  & 0.111  &  101 &  0.084  &  102   & 0.077    &  \\
Sd     &  8    &  52  & 0.062  &   52  &  0.043  &   52     &  0.039   &  \\
Sdm  &  9    &   4   & 0.005  &    4   &  0.003  &   4       &    0.003 &  \\
Sm    &  10  & 4     & 0.005  &    4   &  0.003  &   4       &  0.003   &  \\
\hline 
E-S0   &     &    98  &  0.117  & 158  &  0.131 &  191  &  0.145  & \\
Sa-Sd &     &  708  & 0.848  &  968 &  0.805 &  1035 & 0.785  & \\
Sb-Sc &     &  439  & 0.526  &  633 &  0.526 &   679  &  0.515 & \\

\enddata

\end{deluxetable}

\clearpage

\begin{deluxetable}{lllllll} 
\tablecolumns{7}
\tablewidth{0pc} 
\tablecaption{Morphological code numbers for the UNAM-KIAS sample of Isolated Galaxies.
\label{tbl-1}}
\tablehead{
\colhead{} ID   & SDSS ID  &  T  &  Bar   &  Ring   &}
\startdata

1   &   SDSS J072246.73+413929.6     &     4     &     2     &     1     &     \\
2   &   SDSS J072333.23+412605.5     &     5     &     0     &     1     &     \\
3   &   SDSS J072635.39+431746.8     &     -5     &     0     &     0     &     \\
4   &   SDSS J072642.64+391724.5     &     4     &     1     &     0     &     \\
5   &   SDSS J072719.37+442538.3     &     1     &     0     &     0     &     \\
6   &   SDSS J072954.29+372706.3     &     5     &     0     &     1     &     \\
7   &   SDSS J073054.71+390110.1     &     5     &     0     &     1     &     \\
8   &   SDSS J073548.83+401938.2     &     -     &     0     &     0     &     \\
9   &   SDSS J073901.44+315452.9     &     -2     &     2     &     0     &     \\
10   &   SDSS J074001.42+321140.5     &     2     &     0     &     0     &     \\
11   &   SDSS J074018.11+282751.4     &     2     &     0     &     0     &     \\
12   &   SDSS J074022.74+231629.9     &     -5     &     0     &     0     &     \\
13   &   SDSS J074109.47+492347.0     &     5     &     1     &     0     &     \\
14   &   SDSS J074112.48+424457.7     &     1     &     0     &     0     &     \\
15   &   SDSS J074127.73+294009.0     &     3     &     1     &     1     &     \\
16   &   SDSS J074158.62+231035.0     &     -3     &     0     &     0     &     \\
17   &   SDSS J074232.38+491127.9     &     7     &     0     &     0     &     \\
18   &   SDSS J074252.09+220645.7     &     5     &     0     &     1     &     \\
19   &   SDSS J074330.38+225549.9     &     5     &     0     &     0     &     \\
20   &   SDSS J074403.33+330438.6     &     4     &     1     &     0     &     \\

\enddata

\end{deluxetable}

\clearpage

\begin{figure}[hbt]
\epsscale{0.65}
\plotone{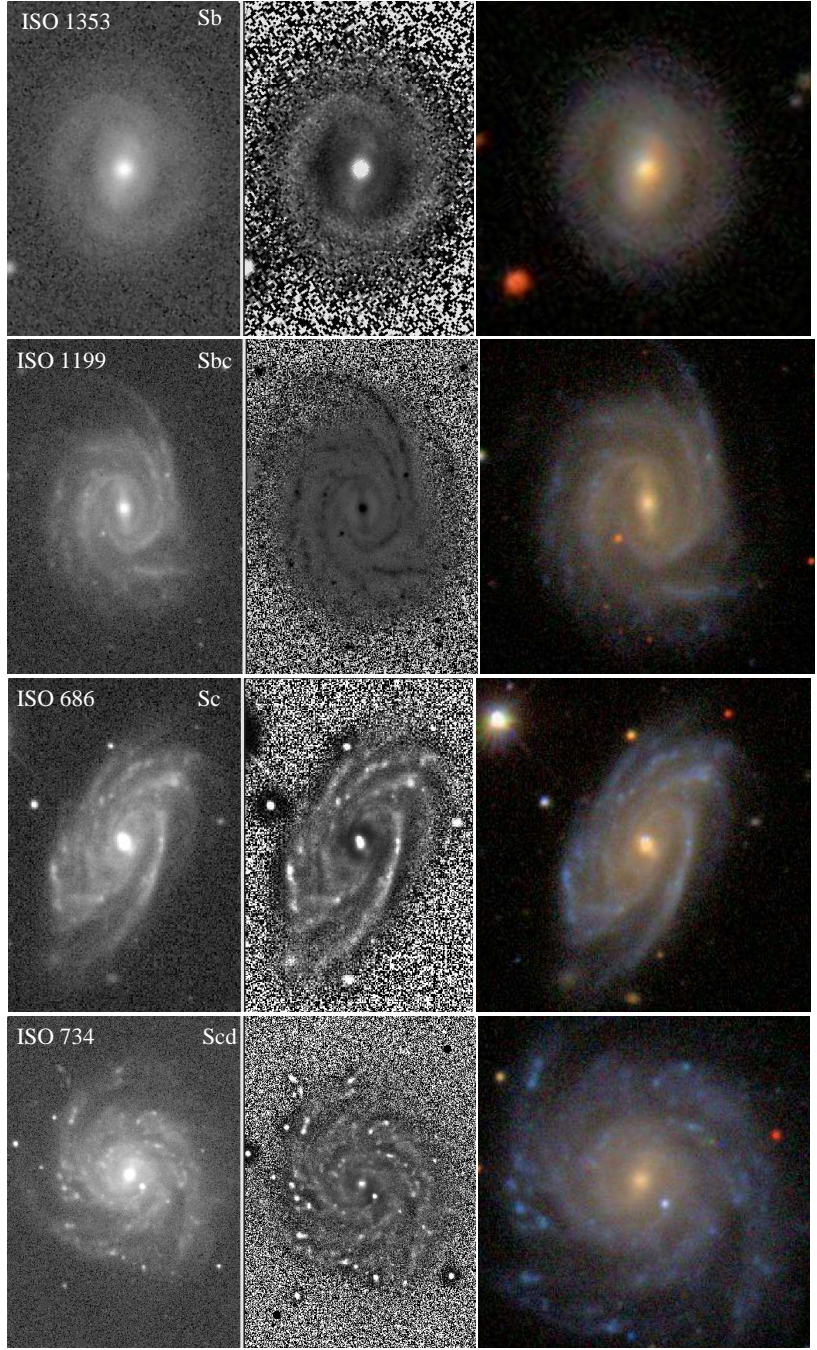}
\caption{Image procedures and Sb, Sbc, Sc and Scd  morphological prototypes  in the UNAM-KIAS isolated galaxy sample.}
\label{mosaico1}
\end{figure} 

\clearpage

\begin{figure}[hbt]
\plotone{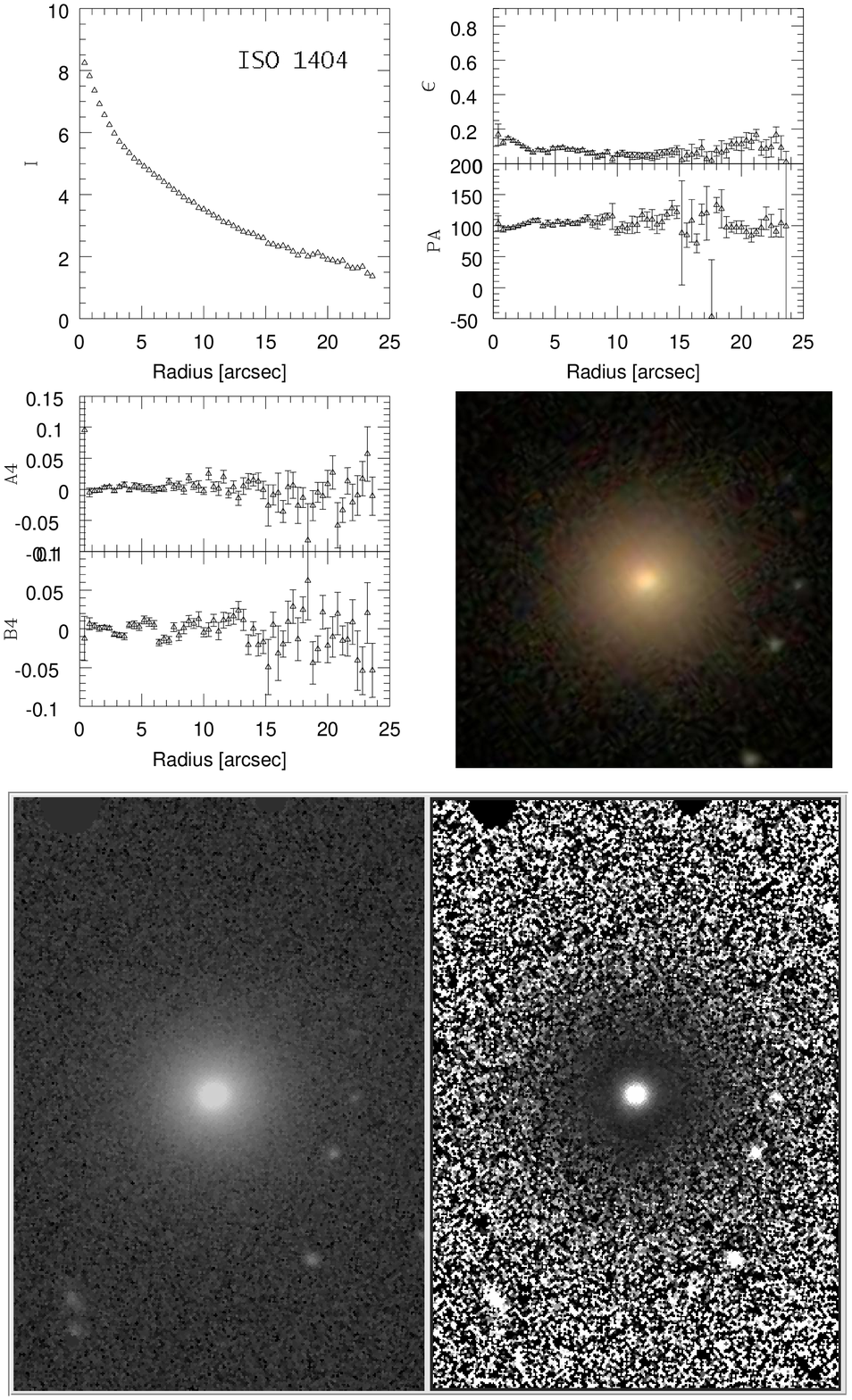}
\caption{ Image procedures and geometric surface photometry profiles for an E morphological prototype in the UNAM-KIAS isolated galaxy sample .}
\label{mosaico1}
\end{figure} 

\clearpage

\begin{figure}[hbt]
\plotone{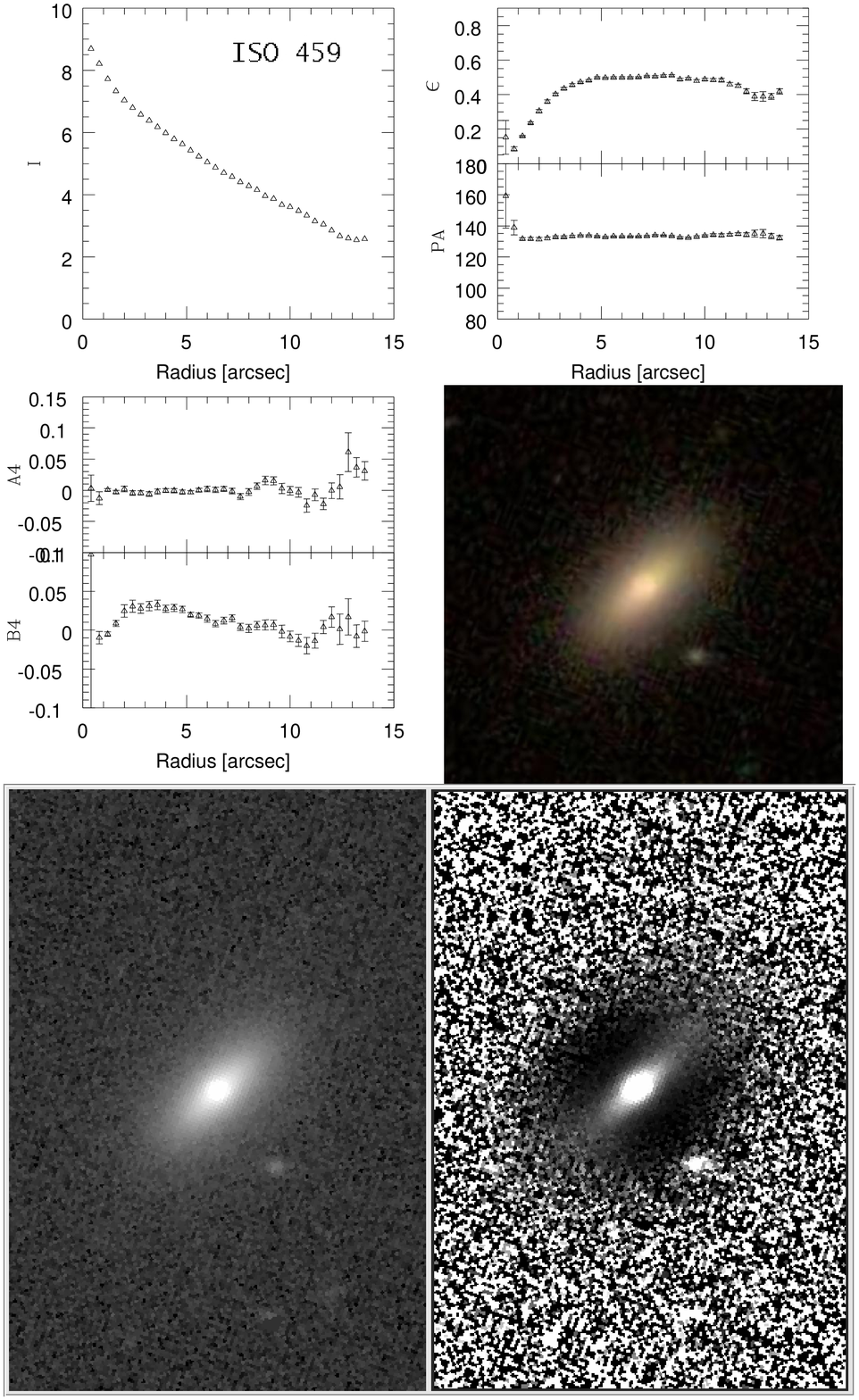}
\caption{Image procedures and geometric surface photometry profiles for an S0 morphological prototype in the UNAM-KIAS isolated galaxy sample .}
\label{mosaico1}
\end{figure} 

\clearpage

\begin{figure}[hbt]
\plotone{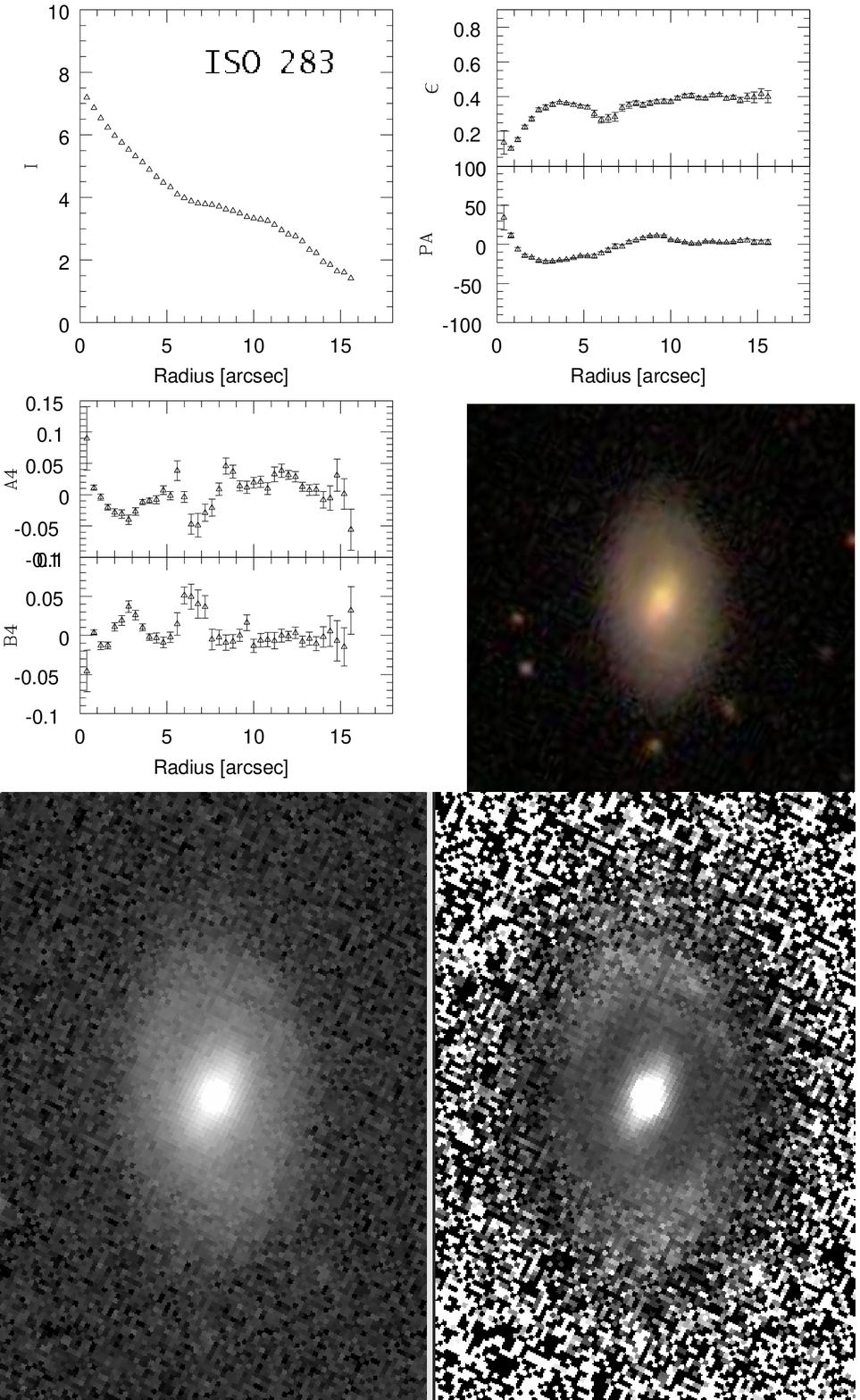}
\caption{Image procedures and geometric surface photometry profiles for an Sa morphological prototype in the UNAM-KIAS isolated galaxy sample.}
\label{mosaico1}
\end{figure} 

\clearpage

\begin{figure}
\epsscale{1}
\plotone{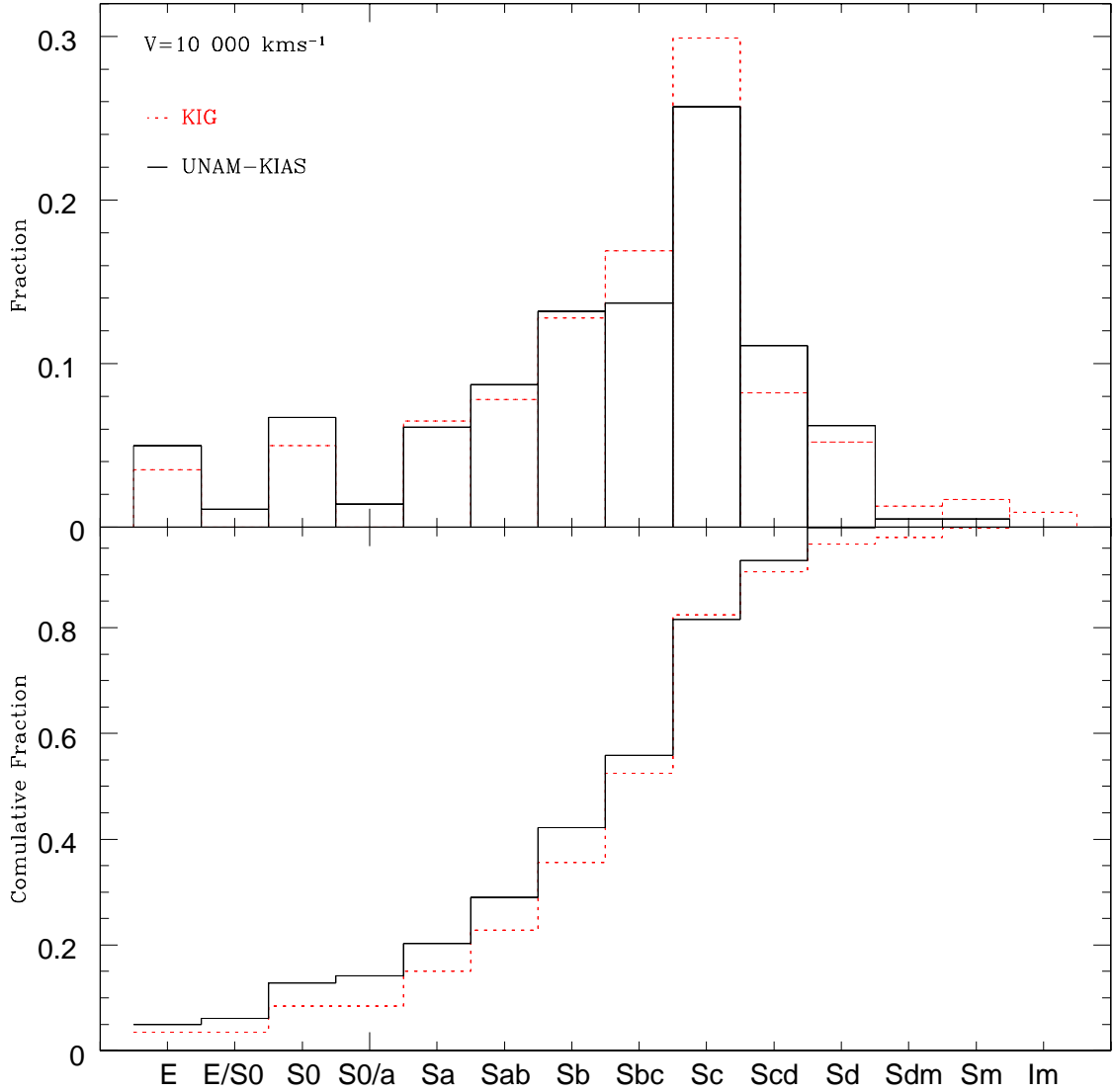}
\caption{Upper panel: The morphological fraction in the UNAM-KIAS (DR5) isolated galaxy sample (black histogram) up to 
$v < 10000 km s^{-1}$ and for the CIG isolated galaxies in common to the SDSS (DR6) database (red histogram). 
 Lower panel: The corresponding cumulative fraction of morphological types.}
\label{mosaico1}
\end{figure}

\clearpage

\begin{figure}
\plotone{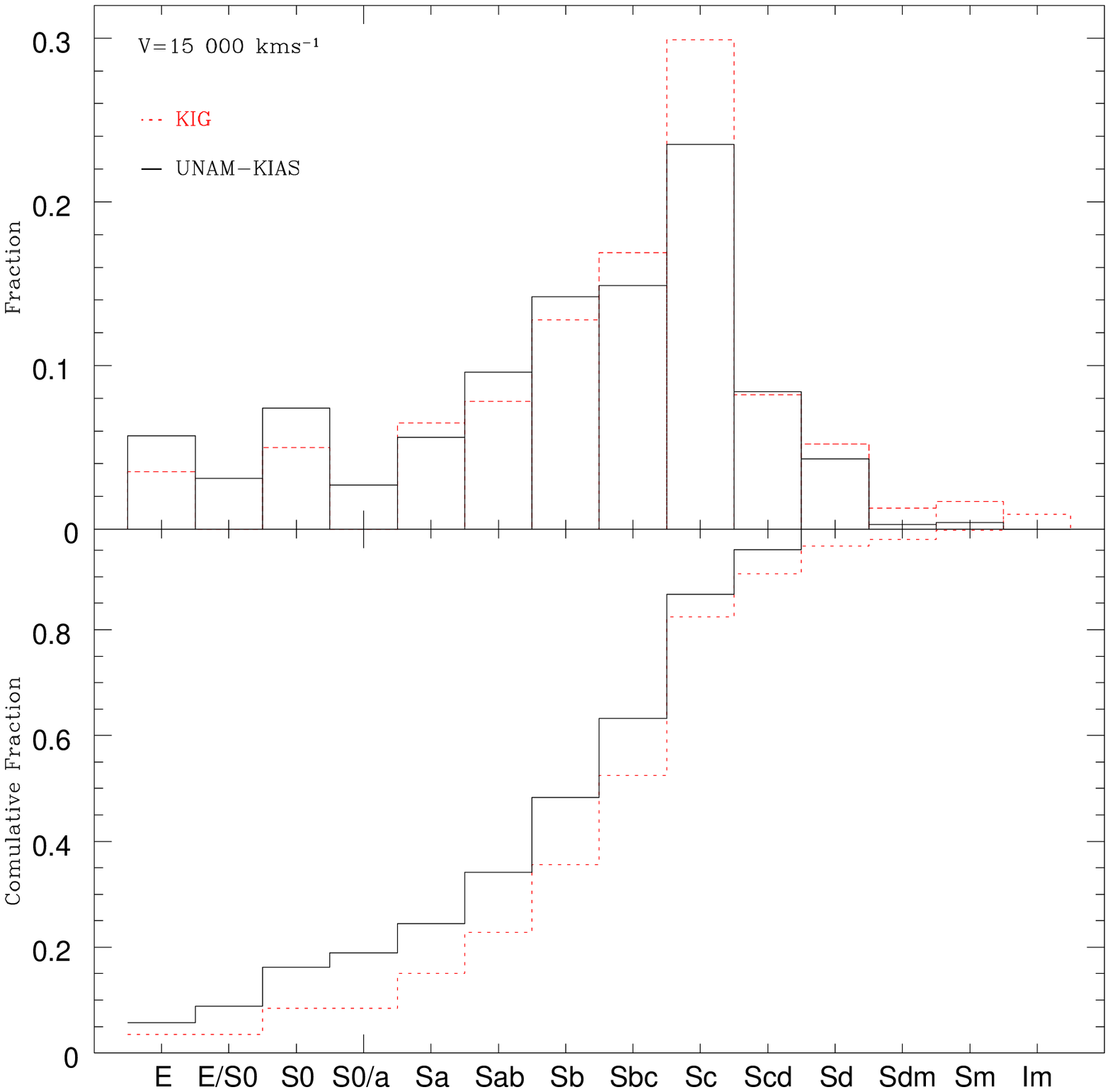}
\caption{Similar to Figure 5 but for the UNAM-KIAS sample up to $v < 15000 km s^{-1}$.}
\label{mosaico1}
\end{figure}

\clearpage

\begin{figure}
\plotone{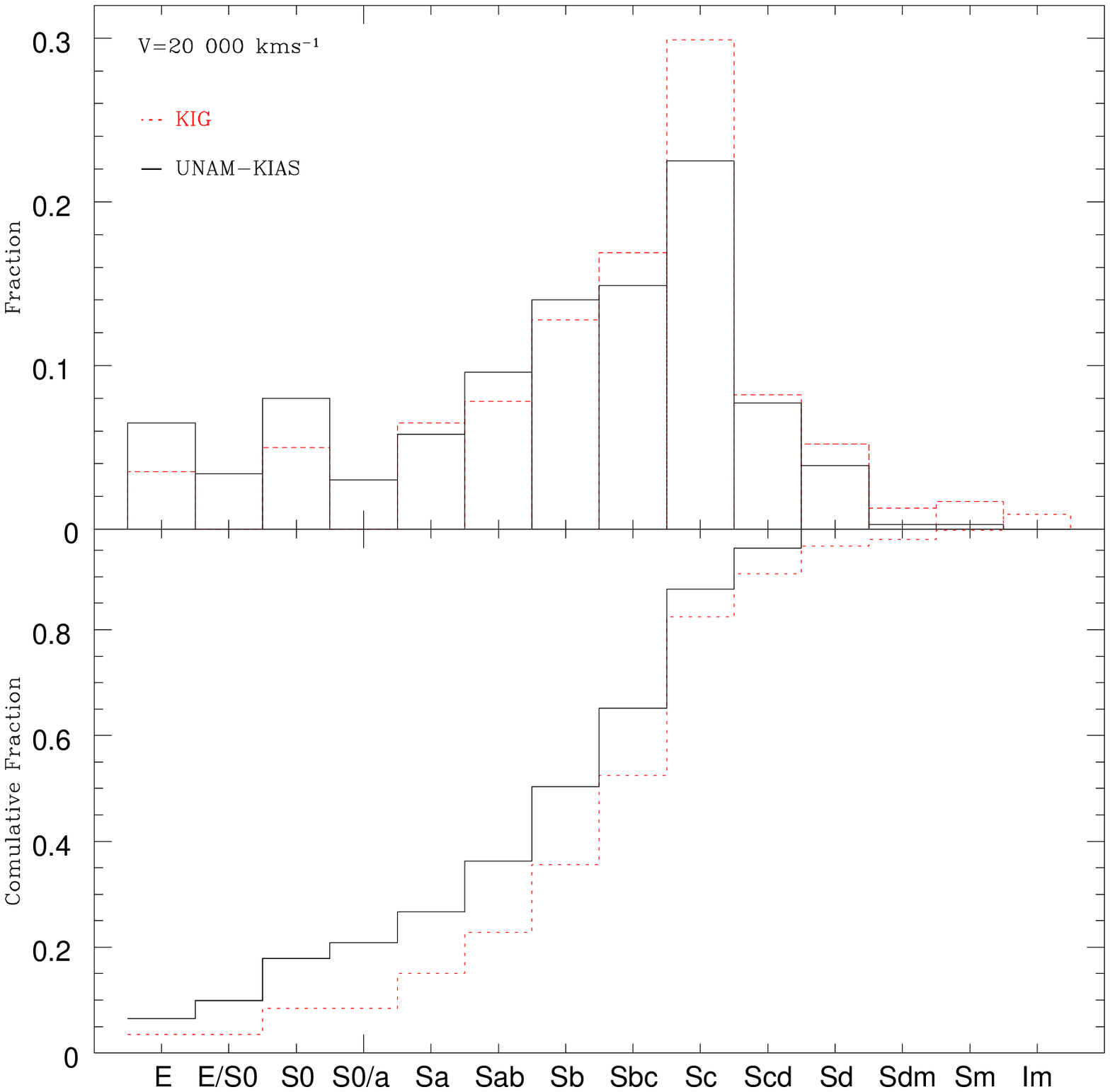}
\caption{Similar to Figure 5 but for the UNAM-KIAS sample up to $v < 20000 km s^{-1}$.}
\label{mosaico1}
\end{figure}

\clearpage

\begin{figure}
\plotone{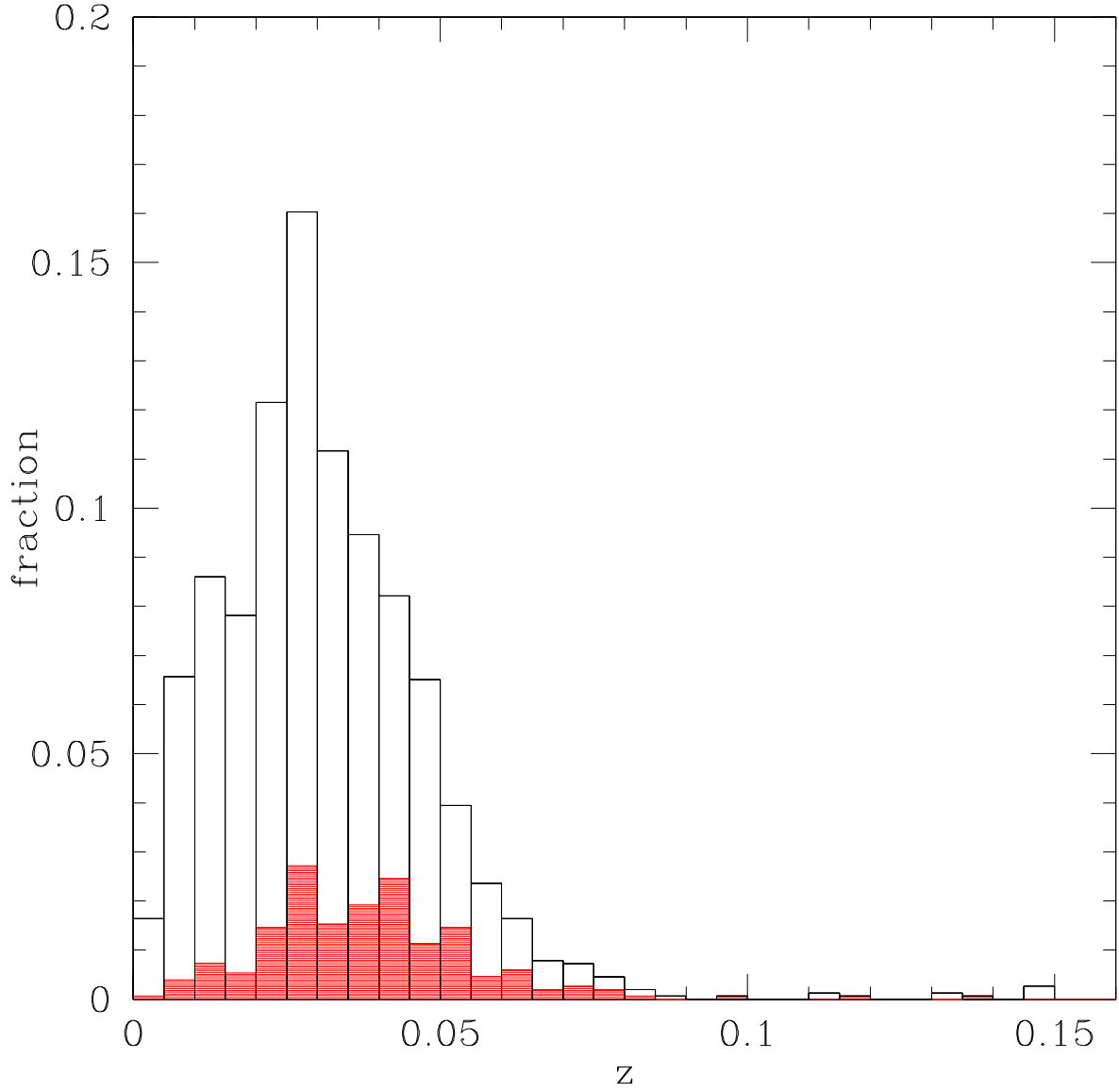}
\caption{The Redshift distribution of the UNAM-KIAS isolated galaxy sample, sorted into E/S0 (red) and Spiral galaxies (black).}
\label{mosaico1}
\end{figure}

\clearpage

\begin{figure}
\plotone{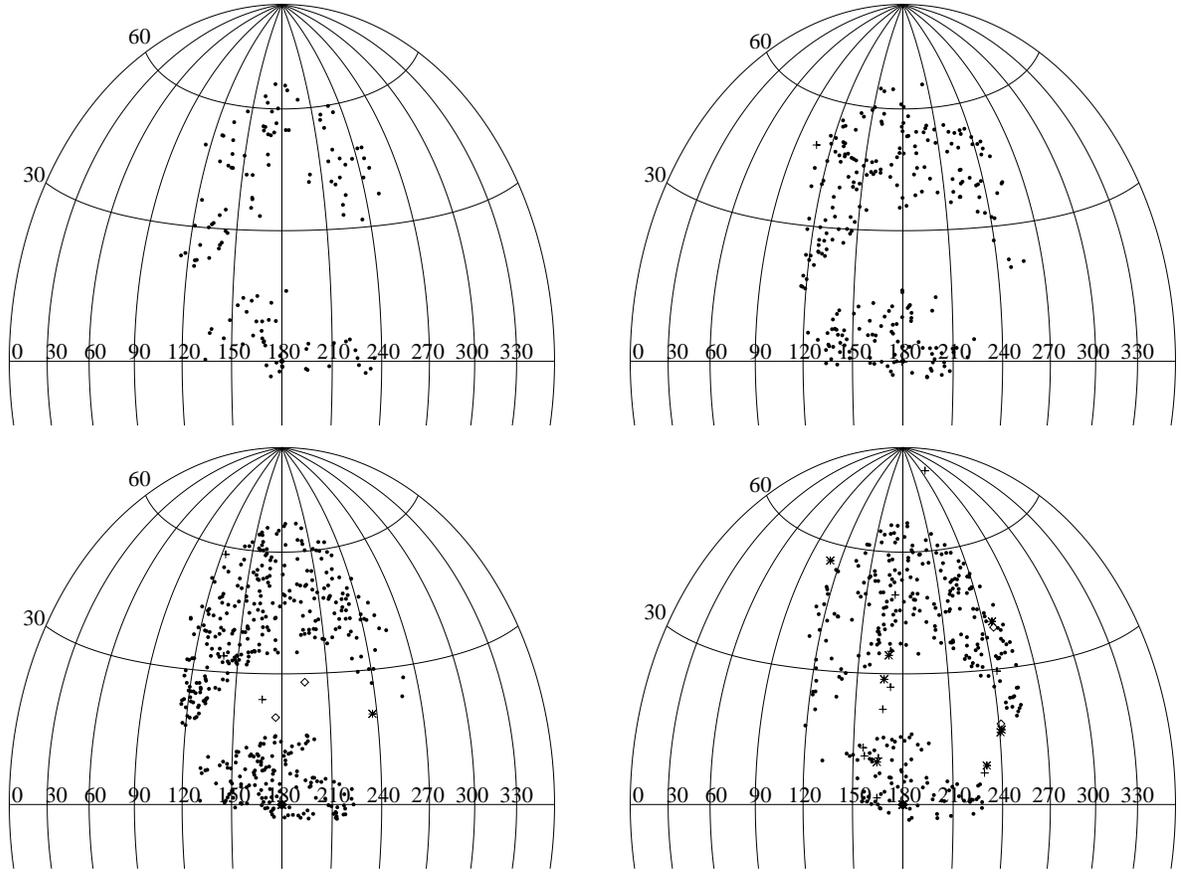}
\caption{Aitoff projections in right ascension and declination showing the distribution of the UNAM-KIAS sample on the sky at 3000 
km s${-1}$ velocity intervals from 0 to 12000 km $s^{-1}$. Abell cluster cores of increasing richness classes from 0 (crosses) 1 (asterisks), 
2 (rhombus) and 3 (triangles) are also indicated.}
\label{mosaico1}
\end{figure} 

\clearpage

\begin{figure}
\plotone{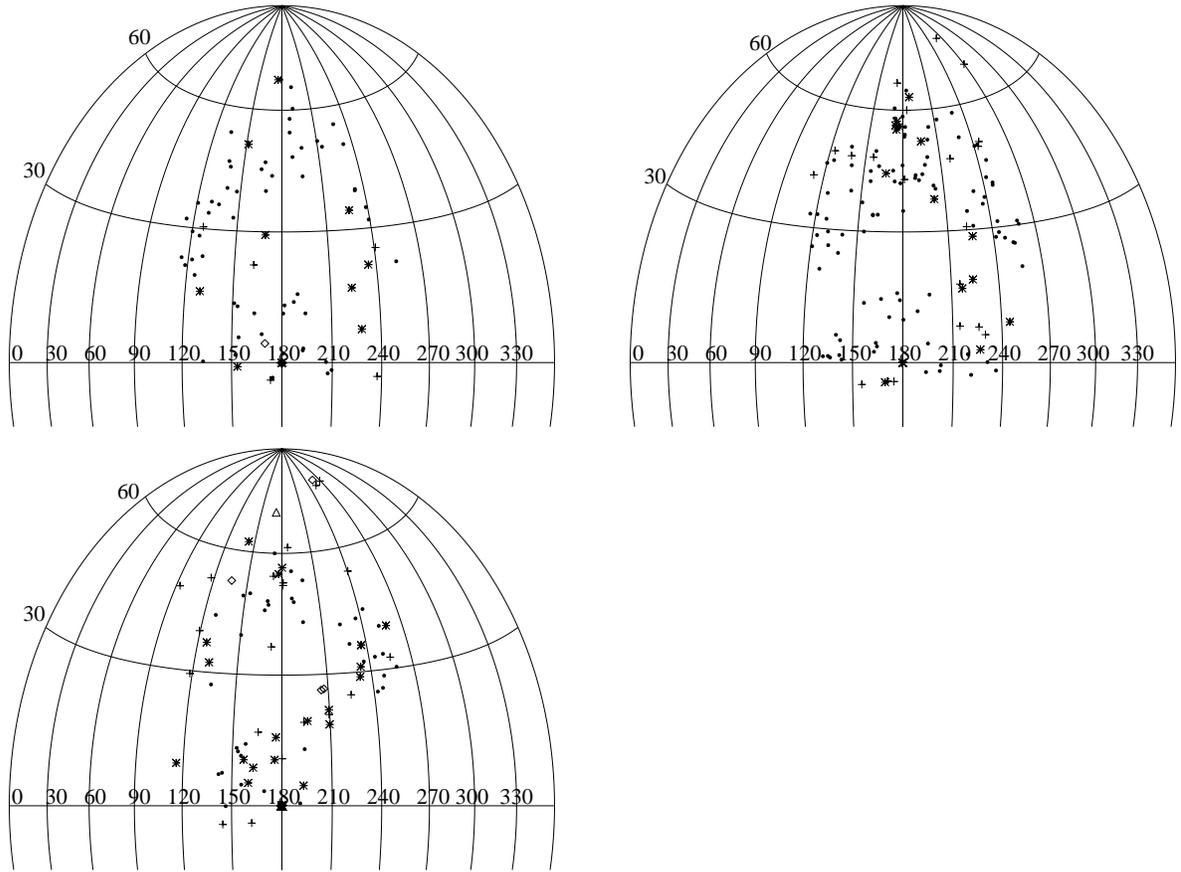}
\caption{Similar to Figure 9 but  from 12000 to 21000 km $s^{-1}$.}
\label{mosaico1}
\end{figure}

\clearpage

\begin{figure}
\plotone{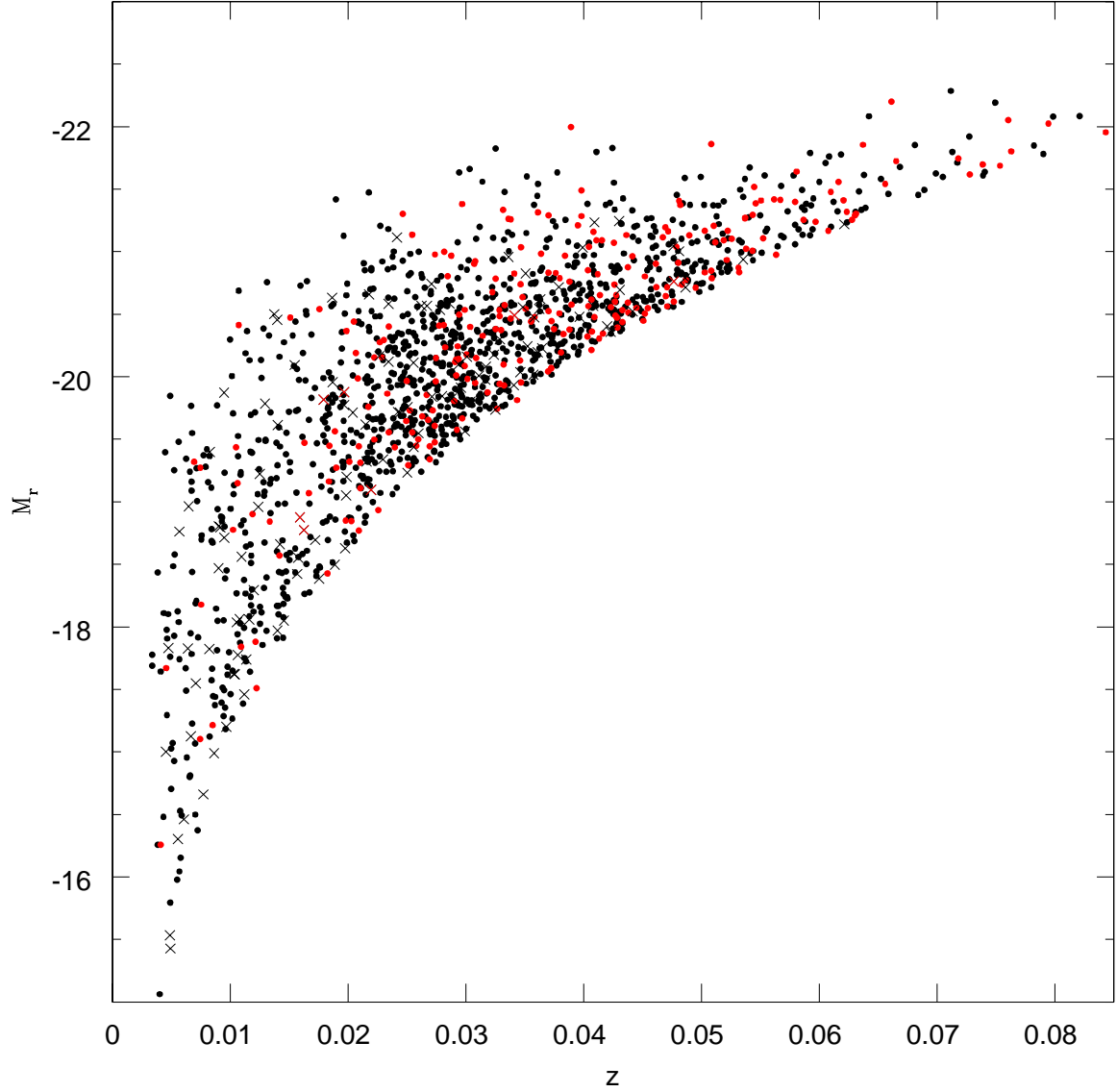}
\caption{The $r$ band Absolute Magnitude vs Redshift Diagram for  the UNAM-KIAS isolated galaxy sample.
E/S0 galaxies are in red; Sa-Sm are in black symbols. Cross symbols indicate galaxies with inclinations greater than $70^{o}$.}
\label{mosaico1}
\end{figure} 

\clearpage

\begin{figure}
\plotone{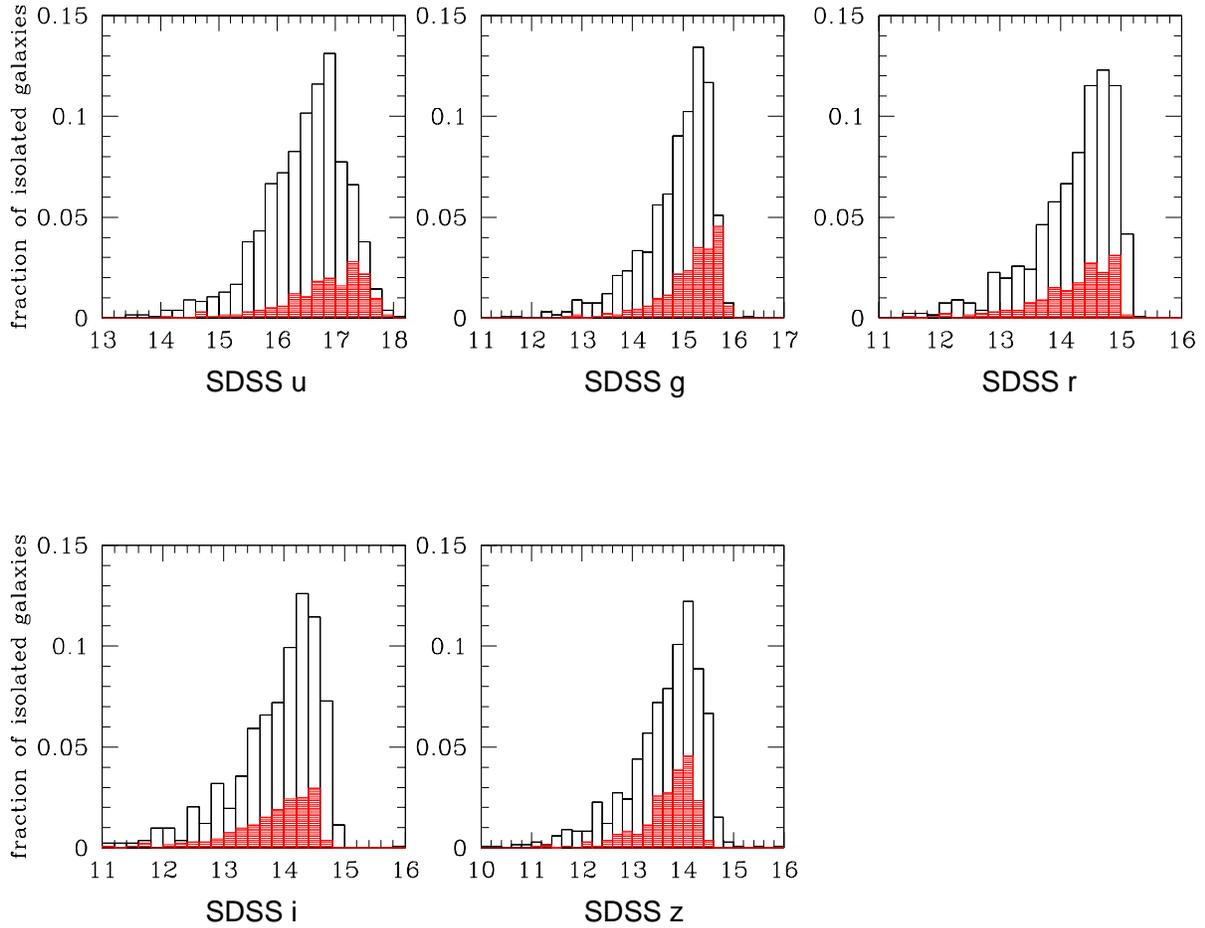}
\caption{Apparent magnitude distributions of the UNAM-KIAS isolated galaxy sample in the $u,g,r,i,z$ filters. Magnitudes are corrected 
for galactic extinction. Red histograms are for E/S0 galaxies. Black histograms are for Sa-Sm galaxies.}
\label{mosaico1}
\end{figure} 
 
\clearpage

\begin{figure}
\plotone{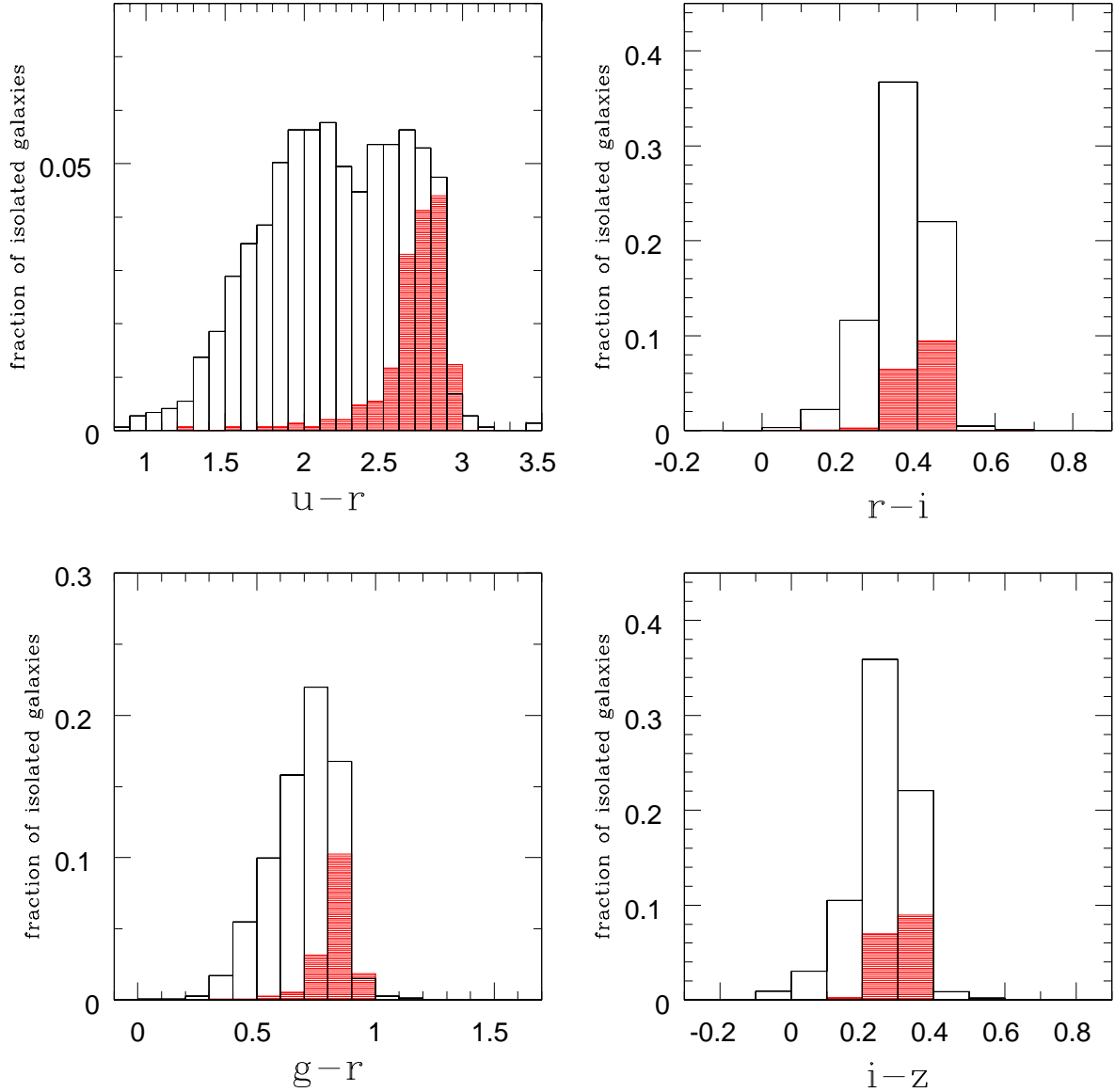}
\caption{($u-r$), ($g-r$), ($r-i$) and ($i-z$) color distributions of the UNAM-KIAS isolated galaxy sample. Colors are corrected 
for galactic extinction. Red histograms are for E/S0 galaxies. Black histograms are for Sa-Sm galaxies.}
\label{mosaico1}
\end{figure} 

\clearpage

\begin{figure}
\plotone{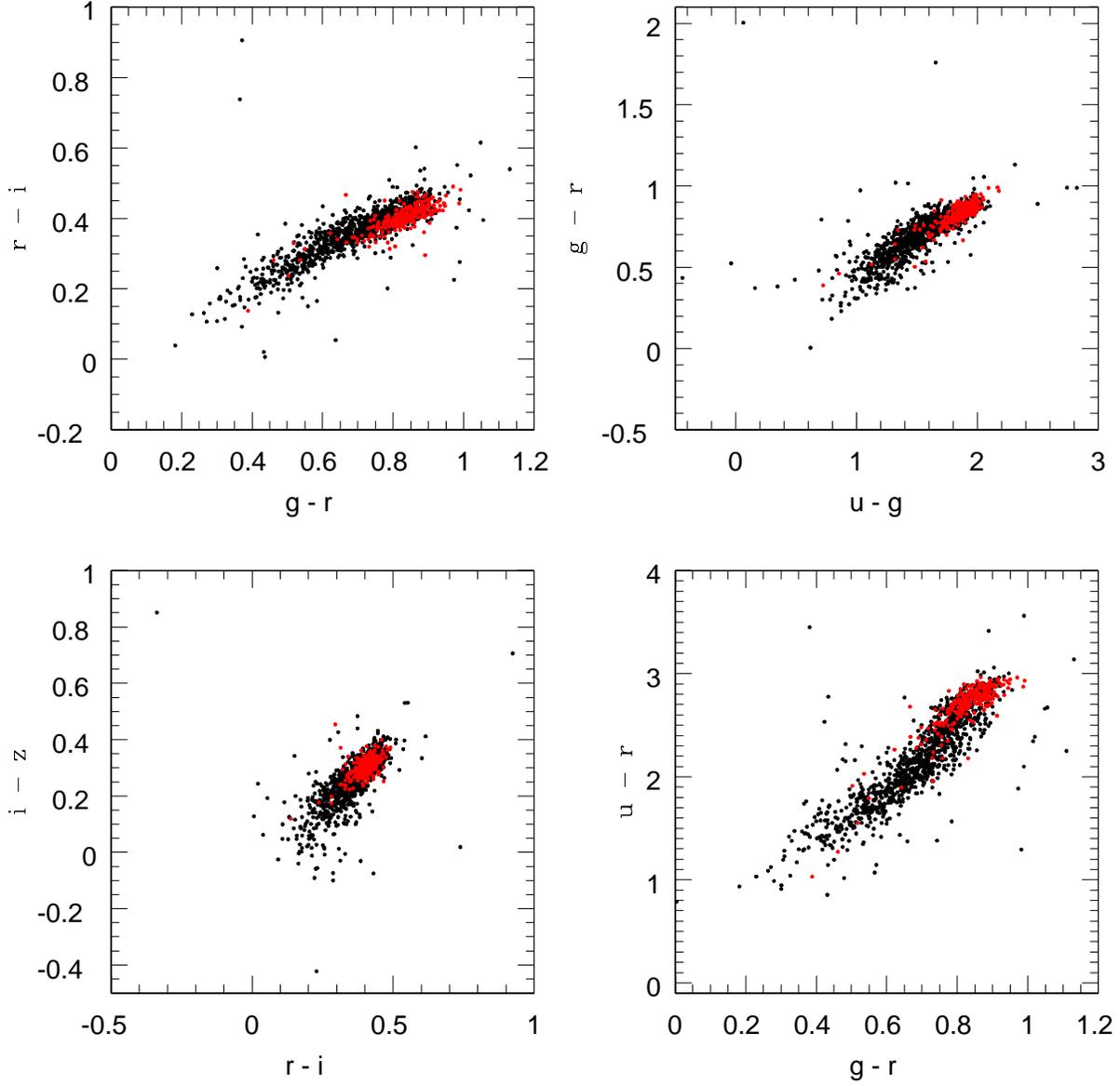}
\caption{($g-r$) vs ($r-i$), ($r-i$) vs ($i-z$), ($u-g$) vs ($g-r$) and ($g-r$) vs ($u-r$) color-color diagrams of the UNAM-KIAS isolated galaxy sample.
E/S0 galaxies are in red; Sa-Sm are in black symbols.}
\label{mosaico1}
\end{figure} 

\clearpage

\begin{figure}
\plotone{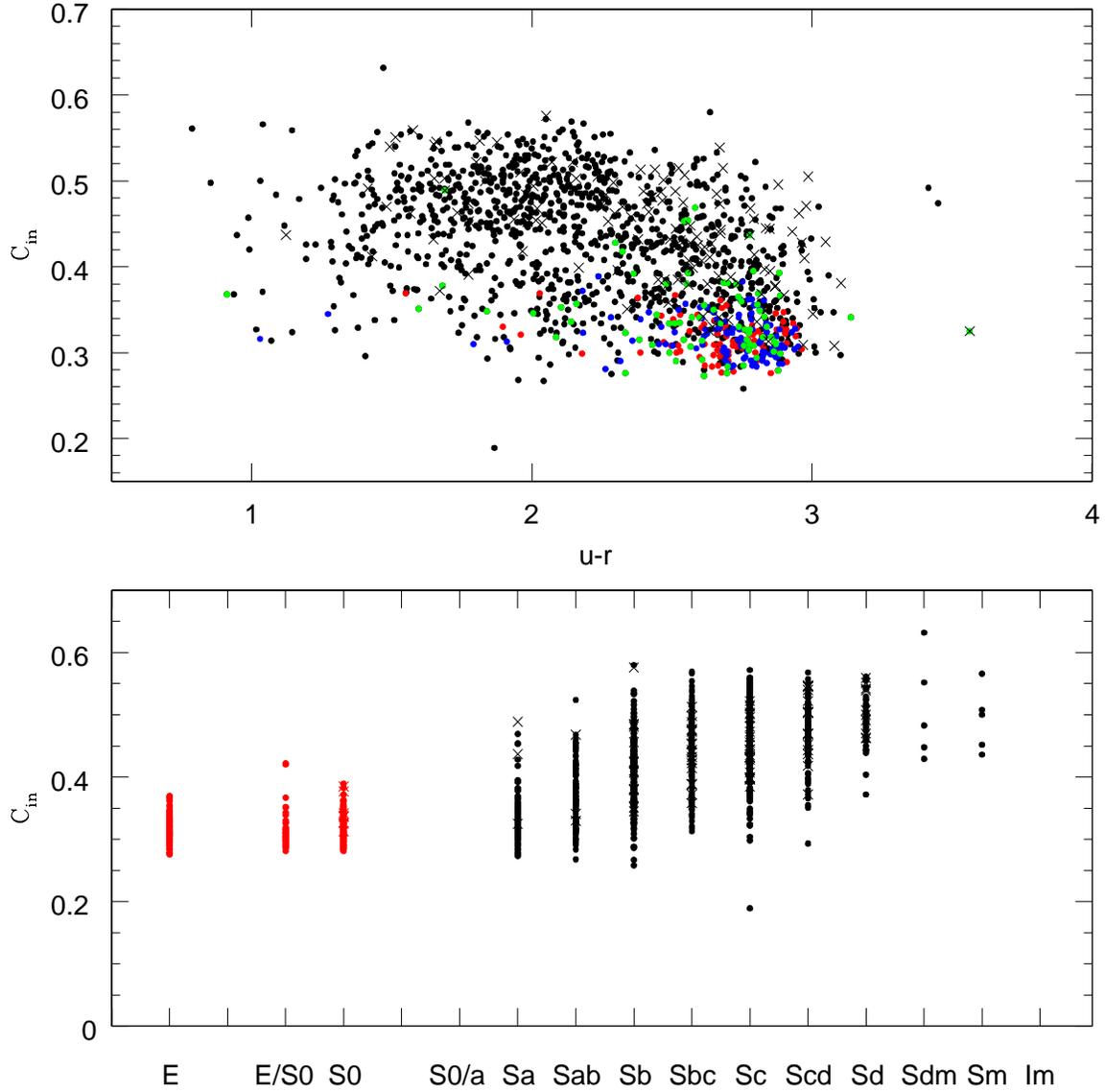}
\caption{Upper panel: Inverse Concentration vs ($u-r$) color diagram. Lower panel: Inverse Concentration vs Morphological type diagram 
for galaxies in the UNAM-KIAS isolated galaxy sample. E/S0 galaxies are in red; Sa-Sm are in black symbols. Cross symbols indicate galaxies with 
inclinations greater than $70^{o}$. Red dots are for E galaxies, blue dots for S0 galaxies and green dots for Sa galaxies. }
\label{mosaico1}
\end{figure}

\clearpage

\begin{figure}
\plotone{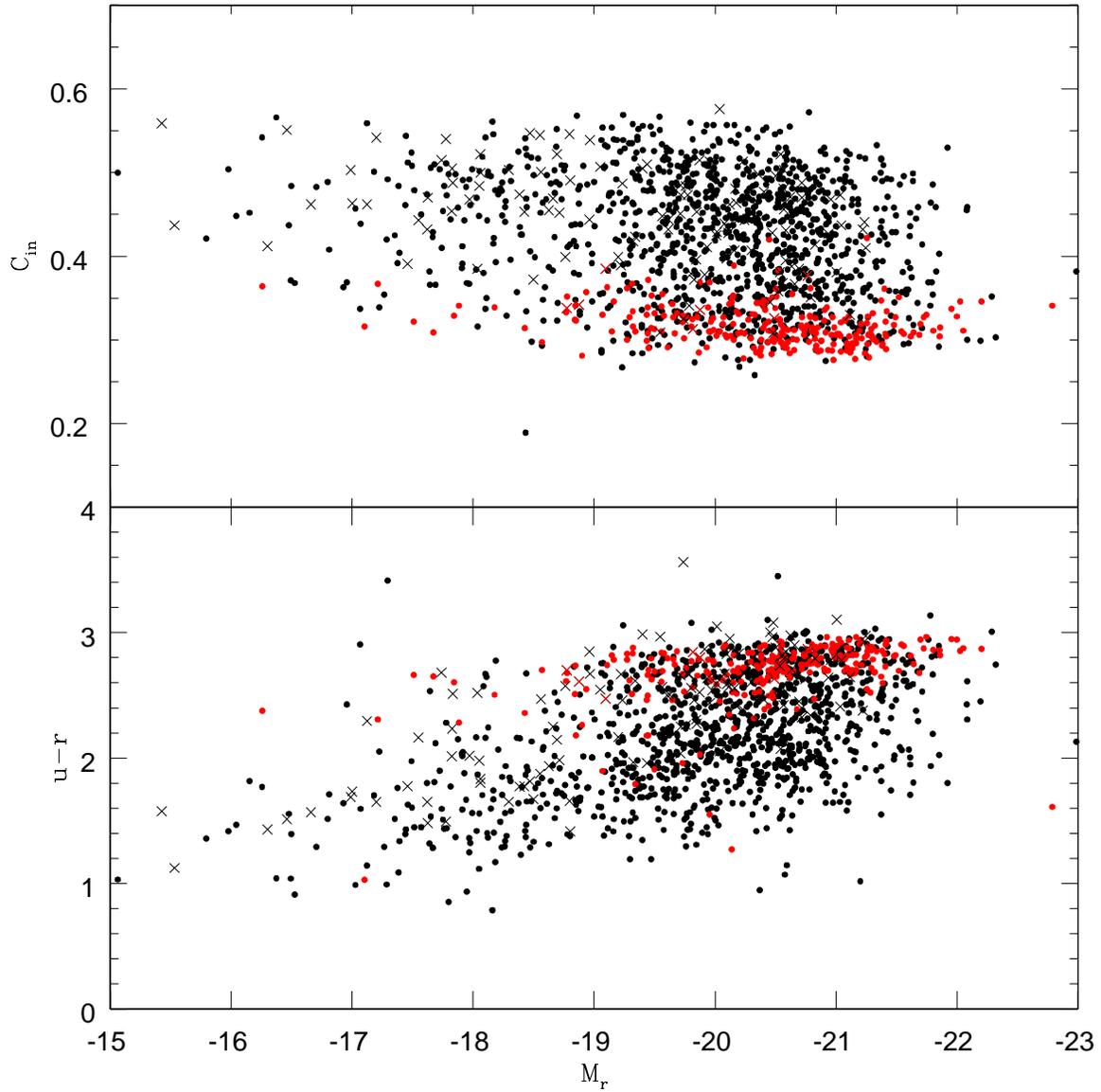}
\caption{Upper panel: Inverted Concentration vs $r$ band Absolute Magnitude diagram. Lower panel: ($u-r$) color vs $r$ band Absolute Magnitude 
diagram  for galaxies in the UNAM-KIAS isolated galaxy sample. E/S0 galaxies are in red; Sa-Sm are in black symbols. Cross symbols indicate galaxies with 
inclinations greater than $70^{o}$.}
\label{mosaico1}
\end{figure} 

\clearpage

\begin{figure}
\plotone{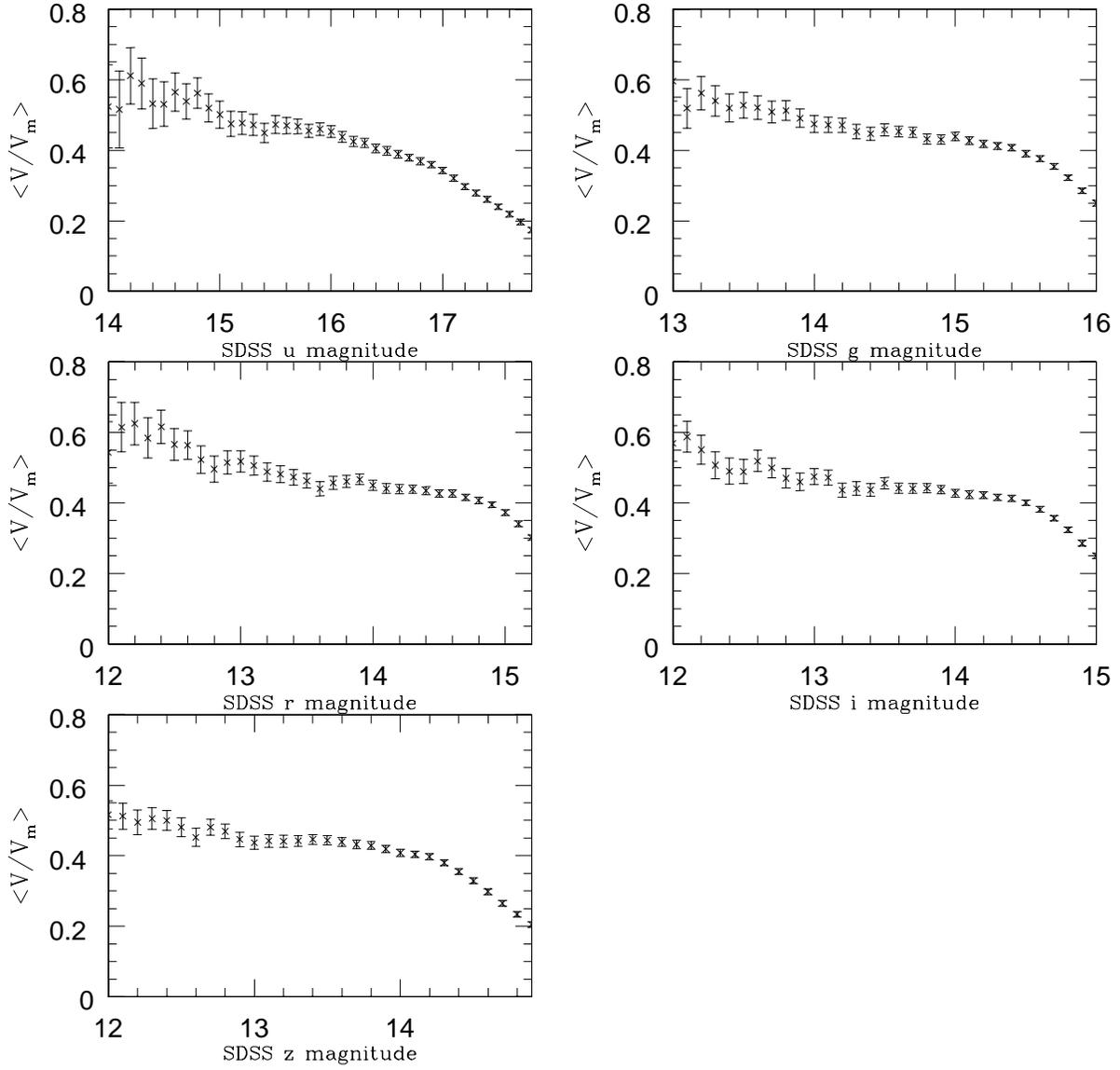}
\caption{ The V/$V_{max}$ test and the completeness of the UNAM-KIAS isolated galaxy sample.}
\label{mosaico1}
\end{figure} 

\clearpage

\end{document}